\documentclass[letterpaper,twocolumn,10pt]{article}
\usepackage{usenix}

\usepackage{tikz}
\usepackage{amsmath}
\usepackage{amssymb}
\usepackage{amsthm}
\usepackage{comment}
\usepackage{booktabs}
\usepackage{graphicx}
\graphicspath{{./images/}}
\usepackage{float}
\usepackage{diagbox}
\usepackage{multirow}
\usepackage{subfigure}

\usepackage{url}

\usepackage{algorithm,algpseudocode}

\usepackage{custom_macros}

\newtheorem{proposition}{Proposition}

\makeatletter
\def\thanks#1{\protected@xdef\@thanks{\@thanks
        \protect\footnotetext{#1}}}
\makeatother

\begin{document}

\date{}

\title{\Large \bf Adversarial Detection Avoidance Attacks: Evaluating the robustness of perceptual hashing-based client-side scanning}

\author{
 \rm{
    Shubham Jain*\thanks{*The first two authors contributed equally and are listed in random order.},
    Ana-Maria Cre\c{t}u*, 
    and Yves-Alexandre de Montjoye$^{\dagger}$\thanks{$^{\dagger}$ Corresponding author at deMontjoye@imperial.ac.uk}} 
\\\\Department of Computing and Data Science Institute, Imperial College London
} 

\maketitle

\begin{abstract}

End-to-end encryption (E2EE) by messaging platforms enable people to securely and privately communicate with one another. Its widespread adoption however raised concerns that illegal content might now be shared undetected. Following the global pushback against key escrow systems, client-side scanning based on perceptual hashing has been recently proposed by tech companies, governments and researchers to detect illegal content in E2EE communications. We here propose the first framework to evaluate the robustness of perceptual hashing-based client-side scanning to detection avoidance attacks and show current systems to not be robust. More specifically, we propose three adversarial attacks--a general black-box attack and two white-box attacks for discrete cosine transform-based algorithms--against perceptual hashing algorithms. In a large-scale evaluation, we show perceptual hashing-based client-side scanning mechanisms to be highly vulnerable to detection avoidance attacks in a black-box setting, with more than 99.9\% of images successfully attacked while preserving the content of the image. We furthermore show our attack to generate diverse perturbations, strongly suggesting that straightforward mitigation strategies would be ineffective. Finally, we show that the larger thresholds necessary to make the attack harder would probably require more than one billion images to be flagged and decrypted daily, raising strong privacy concerns. Taken together, our results shed serious doubts on the robustness of perceptual hashing-based client-side scanning mechanisms currently proposed by governments, organizations, and researchers around the world.\footnote{This is a revised version of the paper published at USENIX Security '22. We now use a semi-automated procedure to remove duplicates from the ImageNet dataset (see Appendix ~\ref{appendix:duplicate-analysis} for details).}
\end{abstract}

\vspace{-0.1cm}
\section{Introduction}
\vspace{-0.1cm}
\begin{figure}[hp]
\centering
\includegraphics[scale=0.41]{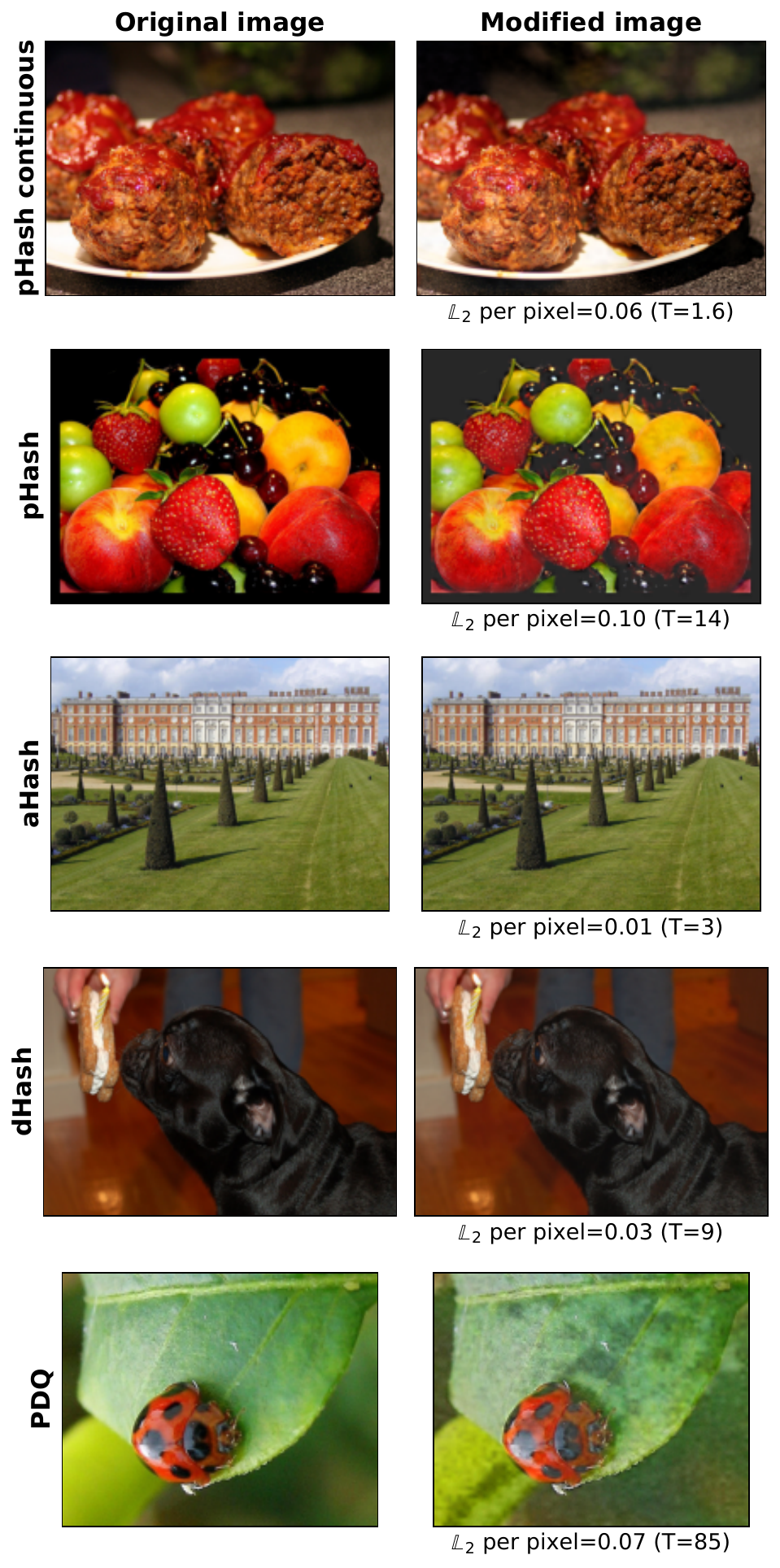}
\caption{Examples of original and modified images for different hashing functions. All four modified images evade detection for the threshold $T$ while maintaining a small $\mathbb{L}_2$ perturbation per pixel and high visual similarity to the original image.}
\label{fig:perturbation-images-paper}  
\end{figure}

More than two billion people across the world use end-to-end encryption (E2EE)-enabled platforms such as Signal and WhatsApp~\cite{signal2021, whatsapp2billion}, exchanging more than 100 billion messages daily on WhatsApp alone~\cite{whatsapp100billion}. E2EE provides a strong privacy protection to users, preventing governments, hackers, and platform providers themselves to access the content of their communications. Governments and organizations have however raised concerns that E2EE is preventing the detection of illegal content~\cite{ncmec2019, international2020e2ee} such as child sexual abuse media and terrorism-related content~\cite{ukcsealaw}. Governments have also recently been discussing bills looking at online safety including the countering of disinformation and misinformation online~\cite{ukonlineharms}.

Following strong concerns from security experts and former national security officials alike~\cite{principles2018key, formernsa2021vice}, key escrow systems providing governments an encryption ``backdoor'' have been abandoned. Instead, client-side scanning has been proposed to detect the sharing of illegal content on E2EE-enabled platforms by tech companies~\cite{applechildsafety}, researchers~\cite{gupta2018whatsapp, reis2020can, mayer2019content}, and policy makers~\cite{eu2020doc}. Here, a signature of a visual media (image, video) would be computed on the user's device and then compared against the database of signatures of known illegal images. If a match is found, the user would e.g. be flagged and/or the unencrypted content automatically shared for further review. Designed to be robust to small changes to the media, as well as transformations like rotation and rescaling, perceptual hashing algorithms would be used to generate the signature. Several variations of this scheme have been proposed. For instance, Apple's recent proposal sends a cryptographic voucher containing the encrypted images and the results of the match. If the number of successful matches reach a predefined threshold, the system is then able to decrypt all the matched images~\cite{applechildsafety}. 

A large literature has developed recently in adversarial machine learning. The input of a machine learning classifier is adversarially modified to fool the model. Slightly modified but visually similar variants of an image can be designed resulting in a misclassification by the model~\cite{sharif2016accessorize, liu2016delving, papernot2017practical}. Some of the most famous examples include an image of panda being misclassified as a gibbon~\cite{Goodfellow_Shlens_Szegedy_2015} or a road works traffic sign to be misclassified as a give way traffic sign~\cite{papernot2017practical}. A wide range of techniques in both white and black-box setup have been developed, including~\cite{madry2018towards} and~\cite{shi2019curls}.

\textbf{Our contribution.} We here propose the first framework to evaluate the robustness of perceptual hashing-based client side scanning to a novel \textit{detection avoidance attack}. Specifically, our framework assesses the threat posed by adversarial attacks aiming to minimally modify an image so as to avoid detection while preserving its content. Our framework also takes into account the diversity of modifications produced by the attack in terms of distances between signatures. The more diverse the perturbation, the harder it is for the system to mitigate by expanding the database with the modified images.

We propose three adversarial attacks against perceptual hashing algorithms, a general black-box attack, inspired by previous work in adversarial ML, e.g., ~\cite{Ilyas_Engstrom_Athalye_Lin_2018, Salimans_Ho_Chen_Sidor_Sutskever_2017, Wierstra_Schaul_Glasmachers_Sun_Peters_Schmidhuber_2014}, and two novel and optimal white-box attacks exploiting the linearity and orthogonality of discrete cosine transform (DCT)-based perceptual hashing algorithms. The attacks are designed to minimally perturb the image while avoiding detection. In particular, they produce a wide range of perturbations preventing easy mitigation strategies such as expanding the database with modified images. 

We evaluate the robustness of perceptual hashing-based client-side scanning mechanisms and show them to be highly vulnerable to detection avoidance attacks. We perform a large-scale extensive evaluation of five commonly used perceptual hashing algorithms: pHash (continuous and discrete)~\cite{Coskun_Sankur_2004, Zauner_2010}, dHash, aHash and PDQ~\cite{facebookpdq} and show that, for all of them, our attack manages avoid detection with modified images very similar to the original. Taken together, our results strongly suggest that perceptual hashing-based client-side scanning is highly vulnerable to small modifications in all scenarios considered, and that simple mitigation strategies like expanding the database would be ineffective. This sheds serious doubts on the robustness of currently proposed client-side scanning mechanisms based on perceptual hashing~\cite{gupta2018whatsapp, eu2020doc, reis2020can, mayer2019content}.

\section{Perceptual hashing-based client-side scanning}
\label{section:background}

\textbf{Perceptual hashing algorithms.} Perceptual hashing algorithms compute a signature of a visual media (e.g., image, video) without having to share it. Perceptual hashes are different from cryptographic hashes in that the former changes gradually as the image changes, while the latter changes significantly as soon as a single pixel changes. Importantly, perceptual hashes are designed to detect instances of a visual media that are visually similar (e.g., a resized version) without being exact copies~\cite{mihccak2001new}. To achieve this, they extract features that remain invariant under small modifications, such as resizing, noise addition, format change and rotation.
For example, pHash and Facebook’s PDQ use the discrete cosine transform (DCT) to extract image features relating to lower frequencies in the DCT output, that remain invariant to image modifications such as blurring, resizing and watermarking~\cite{thornbenchmarking}. 

Formally, a perceptual hashing algorithm $h : \mathcal{I} \rightarrow \mathcal{O}^{l}$ is a deterministic function mapping a multimedia $X \in \mathcal{I}$ to a fixed-size vector representation, the \textit{hash}, usually consisting of bits ($\mathcal{O}=\{0, 1\}$) or real numbers ($\mathcal{O}=\mathbb{R}$). The similarity between two media $X, X' \in \mathcal{I}$ is quantified by computing the \textit{distance} between the hashes according to a metric, henceforth denoted by $d$. For bit-valued hashes, $d$ is usually the Hamming distance, while for real-valued hashes a typical choice for $d$ is the Euclidean distance.

\textbf{Client-side scanning.} Perceptual hashing-based client-side scanning (PH-CSS) for illegal image detection consists of a database $\mathcal{D} = \{X_1, \ldots, X_N\}$ of $N$ images $X_i \in \mathcal{I}, \forall 1 \leq i \leq N$, a perceptual hashing algorithm $h$, a distance $d$, and a threshold $T>0$. Given an image $X \in \mathcal{I}$, the detection system computes the distance between the hash $h(X)$ and the hashes of images in the database $h(X_i)$. The image is flagged if there exists $1 \leq i \leq N$ such that $d(h(X_i), h(X)) \leq T$. We note that our framework is applicable more broadly to visual media (e.g., videos).

The detection system performance can be measured using the false positive and false negative rates\cite{Zauner_2010}. 

\textbf{False positive rate.} For a threshold $T$ and database $\mathcal{D}$, the \textit{false positive rate} $FPR(T, \mathcal{D})$ is defined as the probability that a media visually different from those in the database ($X \not\sim \mathcal{D}$) is detected: 
$$FPR(T, \mathcal{D}) = P(\exists 1 \leq i \leq N \,\,:\,\, d(h(X), h(X_i)) \leq T | X \not\sim \mathcal{D})$$

\textbf{False negative rate.} For a threshold $T$ and database $\mathcal{D}$, the \textit{false negative rate} $FNR(T, \mathcal{D})$ is defined as the probability that a media visually similar to medias in the database ($X \sim \mathcal{D}$) is wrongly rejected:
$$FNR(T, \mathcal{D}) = P(\forall 1 \leq i \leq N \,\,:\,\, d(h(X), h(X_i)) > T | X \sim \mathcal{D})$$

For both the $FPR$ and $FNR$, the randomness is taken over the distribution of media shared by users.

The threshold $T$ modulates the trade-off between the false positive and negative rates. Intuitively, as the threshold $T$ decreases the system will wrongly flag fewer media outside the database. At the same time, it would wrongly reject more media visually similar to those in the database. 

\section{Attack model}
\label{section:attack_model}

\subsection{Detection avoidance attack}
We here propose an adversarial attack against perceptual hashing-based client-side scanning, which we call \textit{detection avoidance attack}. 

\textbf{Attack model.}  We assume that a malicious agent, \textit{the attacker}, is in possession of a image from the database, henceforth denoted \textit{original image} $X \in \mathcal{D}$. The attacker's goal is to minimally modify $X$ into $X'$ such that its content is preserved while avoiding detection. More specifically, the attacker's goal is to modify $X$ via an additive perturbation $\delta$ such that the modified image:  (1) is valid, i.e., $X' = X + \delta \in \mathcal{I}$ meaning that no pixel goes out of bounds, (2) evades detection, i.e., $d(h(X), h(X+\delta)) > T$, where $T$ is the threshold used by the detection system and (3) the perturbation is minimal in terms of visual dissimilarity. We assume that the attacker knows the distance $d$ and the threshold $T$. We revisit these assumptions later.

Assuming that the attacker quantifies the visual similarity between images using a metric $v$ (with smaller values for higher visual similarity), the attacker's goal can alternatively be written as follows.
\begin{align}
    \textrm{Minimize: }\;& v(X, X+\delta) \label{objective}\\
    \textrm{s.t.: } \;& d(h(X), h(X+\delta)) > T \label{constraint:1}\\
    \;& X + \delta \in \mathcal{I} \label{constraint:2}
\end{align}
The objective function quantifies the visual similarity between the original image $X$ and the modified image $X'=X+\delta$. The attacker seeks to minimize this objective in order to preserve the image content. The first constraint requires that the image should avoid detection by not being matched with the original image. However, it is possible the modified image might still be flagged because its hash is close ($\leq T$) to other images in the database.
Finally, the second constraint requires that the modified image should be valid (within the image bounds).

\textbf{Perturbation diversity requirement.} Additionally, the attack should be resistant to simple defenses that the system could implement, such as expanding the database with hashes of modified images. This assumption is realistic as the detection system might gain knowledge of the attack. The attack should therefore produce a wide range of random perturbations that cannot be predicted and added to the database.

\textbf{Attack intuition.} 
We provide three intuitions for why perceptual hashing-based client-side scanning could be vulnerable to the proposed detection avoidance attack. 

First, by design, the perceptual hash of an image changes gradually as the image changes, opening up the possibility to find a minimally modified image $X'$ such that $d(h(X), h(X')) > T$ while $X$ and $X'$ remain visually similar. 

Second, mitigation strategies like increasing the threshold $T$ or expanding the database with hashes of modified images could lead to an increase in the detection system's false positive rate, rendering it unsuitable for the use case. 

Third, the hash space likely contains many valid image hashes that are at least $T$ away from the original image hash. Formally, the $d$-ball of radius $T$ around the hash of the original image $B(X, T) = \{ X' \in \mathcal{I} \,\,:\,\, d(h(X), h(X')) \leq T \}$ is such that its complement $\overline{B(X, T)}$ contains potentially many images $X'$ for typical distances (Hamming, Euclidean) and suitable threshold values. Furthermore, common perceptual hashing algorithms are non-injective, meaning that multiple inputs can lead to the same output. This suggests that it may be possible to obtain several and possibly many different perturbations of the same image such that $d(h(X), h(X')) > T$. 

Finally,  PH-CSS is similar but different from a classification model. It is a threshold-based detection system for matches in a database. This poses unique challenges: first, as the threshold $T$ can be quite large, it is not trivial to see why adversarial perturbations can be found that still preserve the original image content, and second, the detection system could expand the database with hashes of adversarial images produced using our attack, making perturbation diversity a core requirement. 

\subsection{Attacker access to the perceptual hashing algorithm}
We consider two levels of attacker access to the perceptual hashing algorithm.

\textbf{Black-box access.} Unless otherwise specified, we assume that the attacker has black-box access to the perceptual hashing algorithm. In a black-box setting, the attacker can provide an input image $X$ to retrieve its corresponding hash $h(X)$ but does not know how the algorithm works. We believe this assumption to be realistic in the context of perceptual hashing-based client-side scanning context. Most of the proposed implementations of PH-CSS, including Apple's recent proposal, would compute the image hashes on the device~\cite{callas2020thoughts, applechildsafety}. Our black-box assumption is further supported by the fact that Apple's model was made accessible in a recent iOS version, and by subsequent statements by Apple that this was "expected behavior" ~\cite{motherboardapple}.

We further assume that the attacker can retrieve output hashes for as many inputs as needed, without the hashes being uploaded to the server. This can be made possible for example by automating the image upload through the app and creating a honeypot to catch all the requests, read the hash sent in the request and send it back to the attacker. 

\textbf{White-box access.} In some cases, the attacker could have full knowledge of the algorithm used by the detection system. This can for example be the result of reverse-engineering work and the limited number of available perceptual hashing algorithms available. This knowledge consists of the rules or transformations used by the algorithm to map the input to the output hash. In this paper, we develop two novel attacks for DCT-based perceptual hashing algorithms such as the popular pHash and Facebook's PDQ (see Sec. \ref{attack-methodology-white-box}). These attacks provide a strong attack targeting DCT as they are optimal in the sense of minimally modifying the input to evade detection. They also provide theoretical insights into the vulnerability of existing approaches.

\section{Attack methodology}
\label{section:attack_methodology}

We present the methodology for a general black-box attack against perceptual hashing-based client side scanning and two novel attacks against DCT-based hashes.

\subsection{Notation}
\textbf{General.} We consider real-valued images $\mathcal{I}=[0,1]^n$ and define an image as an element $X \in I$. The image size $n$ is equal to the number of pixels in the image. For $1 \leq i \leq n$, we use $X^i$ to denote the $i$-th element of the (flattened) image $X$. Given an image $X \in \mathcal{I}$, the space of valid perturbations is denoted by $\mathcal{I} - X = \{ \delta \in [-1, 1]^n \,\,:\,\, 0 \leq \delta^i + X^i \leq 1, \forall 1 \leq i \leq n \}$ 

\textbf{Distance from original image.} All our attacks aim to modify an original image $X$ into $X'=X+\delta$ such that the distance between their respective hashes is larger than a given threshold $T$.
We denote by $f_{X}: \mathcal{I} \rightarrow \mathcal{O'}$ the function mapping a perturbation $\delta$ to the distance between the hashes of $X'=X+\delta$ and $X$: $f_X(\delta) = d(h(X), h(X+\delta))$. For example, for real-valued hashes $\mathcal{O}' = \mathbb{R}$, while for bit-valued hashes $\mathcal{O}' \subset \mathbb{N}$ (for details see \sref{section:experimental_setup}). For convenience and because each image is attacked separately, we drop the subscript and use $f:=f_X$ when there is no ambiguity. 

\textbf{Visual similarity.} We quantify the visual similarity between original image $X$ and a modified image $X'$ using the $\mathbb{L}_p$ norm of $X-X'$, where $\mathbb{L}_p: X \in \mathbb{R}^n \rightarrow  ||X||_p = \big(\sum\limits_{i=1}^n |X^i|^p \big)^{1/p}$. The Euclidean ($p=2$), $\mathbb{L}_1$ and $\mathbb{L}_{\infty}$ norms are common choices in the adversarial ML literature to measure visual similarity. To compare the visual similarity between original and modified images independently of the image size, we will use the $\mathbb{L}_p$ perturbation per pixel, defined as $\mathbb{L}_{p, \text{pixel}}: X \in \mathbb{R}^n \rightarrow  \big(\frac{1}{n} \sum\limits_{i=1}^n |X^i|^p \big)^{1/p}$.

\subsection{Black-box attack}\label{attack-methodology-black-box}

\textbf{Approach.} Our black-box attack attempts to maximize the distance $f(\delta)$ between the hashes of original and modified images $X$ and $X'=X+\delta$ while ensuring the perturbation $||\delta||_p$ is smaller than a fixed constant $\epsilon$. We take this approach instead of directly minimizing the perturbation under the constraint that $f(\delta) > T$ as we found it impractical to enforce the constraint on $f$ in a black-box setting.

More specifically our attack seeks a solution to the following optimization problem:
\begin{align}
    \textrm{Find: } & \max\limits_{\delta} \min(T, f(\delta)) \\
    \textrm{s.t.: } &  ||\delta||_p \leq \epsilon\\
                    & \delta \in \mathcal{I} - X
\end{align}  

The objective is either $T$ or $f(\delta)$; when it is equal to the former, the program stops, while when it is equal to the latter, i.e., $f(\delta) \leq T$, we perform gradient ascent in search for a better solution. The first constraint requires that the perturbation's $\mathbb{L}_p$ norm does not exceed $\epsilon$. The second constraint requires that the modified image remains valid.

To ensure that the perturbation is as small as possible, we start with a very small admissible perturbation $\epsilon$ and gradually increase it when the above program fails to find a solution within a reasonable amount of steps.

\textbf{Gradient estimation.} We use zero-order gradient estimation via Natural Evolutionary Strategies (NES) ~\cite{Salimans_Ho_Chen_Sidor_Sutskever_2017, Wierstra_Schaul_Glasmachers_Sun_Peters_Schmidhuber_2014} to estimate the gradient of the distance function $f$ with respect to the perturbation $\delta$ and perform gradient ascent. This is a popular technique for optimization under black-box assumptions and has already been used to obtain adversarial perturbations against image classification models~\cite{Ilyas_Engstrom_Athalye_Lin_2018}.

The NES-based strategy for gradient estimation can be described as a special case of estimation using finite-differences on a random Gaussian basis. We estimate the gradient of $f$ using the following equation: 
\begin{equation}\label{equation:gradient_estimate}
    \nabla_{\delta}{E[f(\delta)]} \approx \frac{1}{\sigma d}\sum_{i=1}^{d'}\delta_if(\delta +  \sigma\theta_i)
\end{equation}

where $\theta_i\sim N(0, I_n), 1 \leq i \leq d'$ are samples from a standard multivariate normal distribution over $\mathbb{R}^n$. Nesterov and Spokoiny~\cite{nesterov2017random} showed through theoretical analysis that the number of samples $d'$ required to estimate the gradient scales linearly with the input dimension $n$.

Alg. \ref{algorithm:gradient_estimation} details the black-box gradient estimation procedure. To reduce the variance of our estimate, we use antithetic sampling: we sample Gaussian noise $\theta_i$ for $i \in \{1,...,\frac{d'}{2}\}$ and set $\theta_j = -\theta_{d'-j+1}$ for $j \in \{(\frac{d'}{2} + 1),...,d'\}$ with $d'$ an even number (line 3). This optimization has been empirically shown to improve performance of NES~\cite{Salimans_Ho_Chen_Sidor_Sutskever_2017}. To satisfy the image bound constraints for $X + \delta + \sigma \theta$ which will be given as input to the hashing function (in order to compute $f(\delta + \sigma \theta)$), we replace each noise sample with a clipped version (line 5).

\begin{algorithm}[t]
\caption{\textsc{Grad}: Gradient estimation for $f(\delta)$}
\label{algorithm:gradient_estimation}
    \begin{algorithmic}[1]
        \Inputs{
        $f$: Function whose gradient is to be estimated. \\
        $X$: Original image to be attacked, of size $n$. \\
        $\delta$: Point at which the gradient is to be estimated. \\
        $d'$: Number of Gaussian samples (should be even). \\
        $\sigma$: Scaling factor for Gaussian samples $\sim N(0, I_n)$.\\
        }
        
        \Output{
        $grad$: $\nabla_{\delta}{E[f(\delta)]}$, estimate of the gradient of $f(\delta)$
        }
    
        \Initialize{
        $\theta_i \gets N(0,I_n)$, for $i \in \{1, ...,\frac{d'}{2}\}$\\
        $\theta_i \gets -\theta_{d'-i+1}$, for $i \in \{(\frac{d'}{2} + 1), ...,d'\}$\\
        $grad \gets 0$
        }
    
        \For{$i = 1$ to $d'$}
            \State{
            $\theta_{i}' \gets \max ( \min (1, X + \delta + \sigma \theta), 0)  - X - \delta$}
            \State{
            $grad \gets grad + f(\delta + \theta_i') * \theta_i'* \frac{1}{{\sigma}d'}$
            }
        \EndFor
    \end{algorithmic}
\end{algorithm}

\textbf{Perturbation update.}
Alg. ~\ref{algorithm:update_delta} details the perturbation update.
At each step $t$, we use the sign of the estimated gradient, sign$(grad)$ along with momentum $\mu$ to update $\delta_{t}$ (lines 3-4). Sign gradients have been previously used by Goodfellow et al. to develop adversarial perturbations against image classification models~\cite{Goodfellow_Shlens_Szegedy_2015}. We clip the resulting image $X + \delta_{t+1}$ to ensure it is within the image bounds (line 5), and we enforce the visual similarity constraint on the updated $\delta_{t+1}$ by projecting it onto the ball of $\mathbb{L}_p$ norm $\epsilon$ (lines 6-7).

\begin{algorithm}[t]
\caption{\textsc{Update}: Estimating $\delta_{t+1}$ via gradient ascent}
\label{algorithm:update_delta}
    \begin{algorithmic}[1]
        \Inputs{
        $\delta_{t}$: Perturbation after the $t^{th}$ iteration.\\
        $grad_{t}$: $\nabla_{\delta}{E[f(\delta_t)]}$, estimate of the gradient of $f(\delta_t)$.\\
        $prev\_grad_{t}$: Weighted sum of previous gradients.\\
        $X$: Original image to be attacked.\\
        $\epsilon$: Upper bound on $||\delta||_p$.\\
        $\mu$: Momentum parameter.\\
        $\eta$: Step size to update $\delta$ in each iteration.
        }
        
        \Output{
        $\delta_{t+1}$: Perturbation after the $({t+1})^{th}$ iteration.\\
        $prev\_grad_{t+1}$: Updated weighted sum of gradients.
        }

        \State{$grad_t \gets \mu * prev\_grad_t + (1 - \mu) * grad_t$}
        \State{$\delta_{t+1} \gets \delta_t + \eta * \textrm{sign}(grad_t)$}
        \State{$\delta_{t+1} \gets \min(\max(X + \delta_{t+1}, 0), 1) - X$}
        \If {$||\delta_{t+1}||_p > \epsilon$}
        \State{$\delta_{t+1} \gets \delta_{t+1} * \epsilon / ||\delta_{t+1}||_p$}
        \EndIf
        \State{$prev\_grad_t \gets grad_t$}
    \end{algorithmic}
\end{algorithm}

\textbf{Perturbation bound.} To ensure that the visual similarity is comparable across images of different sizes, the bound on the perturbation norm ($\epsilon$) is derived from a constant independent of the image size (denoted $\epsilon_{\text{norm}}$) for each image attacked. Perturbing each pixel by 1\% in the same image but with different sizes would result in a visually similar perturbation but different values of $||\delta||_p$. In the attack, we increment $\epsilon_{\text{norm}}$ starting from small values, and the corresponding actual perturbation bound used in the attack (dependent on the image size) will be $\epsilon = \epsilon_{\text{norm}} * n^{\frac{1}{p}}$.

\textbf{Making the attack more efficient.} Because the number of samples $d'$ required to estimate the gradient scales linearly with the image size~\cite{nesterov2017random}, we attack a small-dimension resized and grayscaled version of the original image, which we denote by $\bar{X}$ (of size $n' < n$). The attack produces a perturbation $\bar{\delta}$, which we map to a final perturbation $\delta$ in the original space using Alg. \ref{algorithm:inverse-delta}. Our strategy, which we show to work in practice, is informed by three insights: (1) most perceptual hashing algorithms are known to be invariant to grayscaling and resizing of the image~\cite{benchmarkingcontentblockchain}. This means that, for an image $X$, $h(X)$ does not change much even when $X$ is recolored or resized. (2) both grayscaling, i.e., converting a three-channel RGB image to single-channel image, and resizing to a smaller sized image reduce the size of the original image $X$ and (3) optimizing the perturbation for the grayscale image and mapping it to a corresponding perturbation in the RGB space leads to smoother perturbations than directly optimizing for the RGB image.

Alg. \ref{algorithm:inverse-delta} details our algorithm for inverting the perturbation. Using a resize transformation, we map the single-channel perturbation $\bar{\delta}$ (of size $n'$) to a corresponding perturbation of same single-channel size as the original image $X$ ($n/3$), denoted by $\delta_{gray}$. The final pixel value $\delta_c^i$ for channel $c$ is derived from $\delta_{gray}$ such that (1) $\delta_c^i$ is proportional to $X_c^i$, the intensity for pixel $i$ in the original image, and (2) the modified pixel $X_c^i + \delta_c^i$ is within the image bounds.

\begin{algorithm}
\caption{\textsc{InverseDelta}: Calculate perturbation $\delta$ for original image $X$ from perturbation $\bar{\delta}$}
\label{algorithm:inverse-delta}
    \begin{algorithmic}[1]
        \Inputs{
        $X$: Original image to be attacked of size $n$.\\
        $\bar{\delta}$: Optimal perturbation for $\bar{X}$.
        }

        \Output{
        $\delta$: Perturbation for $X$.
        }
        \Initialize{
        $X_{c} \gets$ channel $c$ of $X$, for $c \in \{R, G, B\}$\\
        $\textrm{mean}({X_{c}}) \gets \frac{X_R + X_G + X_B}{3}$\\
        $n_{gray} \gets n/3$
        }
        \State{$\delta_{gray} = \textrm{resize}(\bar{\delta}, n_{gray})$}
        \For{$c \;\textrm{in}\; \{R, G, B\}$}
            \For{$i = 1 \;\textrm{to}\; n_{gray}$}
                \If{$\delta_{gray}^{i} \leq 0$}
                    \State{$\delta_c^i = \delta_{gray}^{i}\frac{X_c^i}{\textrm{mean}({X_{c}^i})}$}
                \Else
                    \State{$\delta_c^i = \delta_{gray}^{i}\frac{1 - X_c^i}{1 - \textrm{mean}({X_{c}^i})}$}
                \EndIf
                \State{$\delta_c^i = min(max(X_c^i + \delta_c^i, 0), 1) - X_c^i$}
            \EndFor
        \EndFor
    \end{algorithmic}
\end{algorithm}

\textbf{Complete attack.} The complete attack for an image and $X$ of size $n$ and threshold $T$ is described in Alg. \ref{algorithm:black-box-attack}. The image-independent perturbation bound is initialized to $\epsilon_0$. We grayscale and resize $X$ to obtain $\bar{X}$ of smaller size $n'$. In each iteration $t$, the algorithm estimates the gradient and performs a gradient update. If the distance $f_{\bar{X}}(\bar{\delta})$ reaches a plateau  as quantified by the finite difference $\Delta f$ (and the number of previous iterations for which $\Delta f \leq 0$), we consider that the current $\epsilon$ is too small to allow for $f(\bar{\delta}) > T$. We thus increase the image-independent perturbation bound $\epsilon_{\text{norm}}$ by a fixed step $\eta_{\epsilon}$ and update $\epsilon$ accordingly. As soon as the distance $f_{\bar{X}}(\bar{\delta})>T$, $\delta$ is computed from $\bar{\delta}$. The algorithm stops if $f_X(\delta)>T$ as well, in which case the attack is considered successful. If after $m$ iterations the distance $f_X(\delta)$ does not exceed $T$, the attack is considered unsuccessful.

\begin{algorithm}[htb!]
\caption{Black-box attack against perceptual hashing algorithms}
\label{algorithm:black-box-attack}
    \begin{algorithmic}[1]
        \Inputs{
        $X$: Original image to be attacked of size $n$.\\
        $h$: Perceptual hashing algorithm to be attacked.\\
        $d$: Distance metric corresponding to perceptual hashing algorithm $h$.\\
        $T$: Threshold to be attacked.\\
        $n' (< n)$: Size of the resized and grayscaled image $\bar{X}$\\
        $\epsilon_0$: Starting value for the maximum perturbation allowed.\\
        $\eta_{\epsilon}$: Step size for $\epsilon$.\\
        $k$: Number of previous iterations used to detect plateauing of $f_{\bar{X}}$.\\
        $\mu$, $\eta$: Parameters required for updating the gradient.\\
        $d'$, $\sigma$: Number of samples and scale for Gaussian samples used in gradient estimation.\\
        $s$: Seed defined by the attacker.\\
        $m$: Maximum number of iterations.\\
        }
        
        \Output{
        $\delta$: Perturbation for original image X.\\
        }
        \Initialize{
        $\bar{X} \gets \textrm{resize}(\textrm{rgb2gray}(X), n')$\\
        $\bar{\delta}_0 \gets 0$\\
        $prev\_grad_{0} \gets 0$\\
        $seed \gets (X, s)$\\
        $\epsilon_{norm} \gets \epsilon_{0}$\\
        $\epsilon \gets \epsilon_{norm}*n^{\frac{1}{p}}$\\
        }

        \State{setseed($seed$)}
        \For{$t= 0 \textrm{ to } m-1$}
            \State{$grad_t \gets \textsc{Grad}(f_{\overline{X}}, \bar{X}, \bar{\delta}_t, d', \sigma)$}
            \State{$\bar{\delta}_{t+1}, prev\_grad_{t+1} \gets \textsc{Update}($\par$\bar{\delta}_{t}, grad_t, prev\_grad_t, \overline{X}, \epsilon, \mu, \eta)$}
            \State{${\Delta}f_t \gets f_{\overline{X}}(\bar{\delta}_{t+1}) - f_{\overline{X}}(\bar{\delta}_t)$}
            \If {$\textrm{Count}([{\Delta}f_{t-k+1},...,{\Delta}f_{t}] \leq 0) > \frac{k}{2}$}
                \State{$\epsilon_{norm} \gets \epsilon_{norm} + \eta_{\epsilon}$}
                \State{$\epsilon \gets \epsilon_{norm}*n'^{\frac{1}{p}}$}
            \EndIf
            \If {$f_{\overline{X}}({\bar{\delta}_{t+1}}) > T$}
                \State{$\delta = \textsc{InverseDelta}(X, \bar{\delta}_{t+1})$}
                \If {$f_{X}(\delta) > T$}
                    \State{\textbf{break}}
                \EndIf
            \EndIf
        \EndFor
    \end{algorithmic}
\end{algorithm}

\textbf{Diversity of perturbations.} For the attack to be resistant against simple mitigation strategies like expanding the database with hashes of the modified image $X'$, the attack should generate multiple random perturbations for an image, resulting in different and unpredictable output hashes. To achieve this objective, we make two modifications to our black-box attack. First, we initialize our perturbation $\bar{\delta}_{0}$ to be a valid random non-zero perturbation such that $||\bar{\delta}_0||_p = \epsilon_{start} * n'^{\frac{1}{p}}$ for some small value of $\epsilon_{start}$. More specifically, we uniformly sample $\bar{\delta}_0$ from a set of valid perturbations, scale it to have an $\mathbb{L}_p$-norm of $\epsilon_{start}$ and clip to limit it within bounds.

\begin{equation}
    \begin{aligned}
        \bar{\delta}_0 &\sim U(-\bar{X}, 1-\bar{X})\\
        \bar{\delta}_0 &= \epsilon_{start} * n'^{\frac{1}{p}} * \frac{\bar{\delta}_0}{||\bar{\delta}_0||_p}\\
        \bar{\delta}_0 &= \min(\max(\bar{\delta}_0, 0), 1)
    \end{aligned}
\end{equation}

Second, we observe that requiring the perturbation to be smaller than the constant used in the default black-box attack can lead to less diverse outputs, even when different starting points are used in the optimization. So we allow for a larger maximum perturbation.

The two algorithm modifications lead to more diverse hashes, at a small cost to the visual similarity.

\textbf{Reproducibility. } In order to ensure our attack results are reproducible, we initialize the attack using both the image $X$ and the attacker-defined seed $s$ (line 4 in Alg.~\ref{algorithm:black-box-attack}). This ensures that the directions $\theta$ being probed for gradient estimation are not the same for all images even when the attacker-defined seed $s$ remains the same. For diversity, using different attacker-defined seeds leads to diverse perturbations for the same image. An attacker can either manually define the seed, or set it based on processor clock to generate a diverse perturbation everytime.

\subsection{White-box attacks for DCT-based hashes}\label{attack-methodology-white-box}

We develop a principled attack to devise a minimum perturbation for DCT-based perceptual hashing algorithms. The Discrete Cosine Transform (DCT)~\cite{Ahmed_Natarajan_Rao_1974} is a very popular image compression algorithm.  DCT maps a discrete signal to linear combinations of the original signal and cosines of different frequencies. The signal can then be compressed by noting that coefficients corresponding to lower frequencies encode the most important features. This property makes DCT suitable for JPEG image compression~\cite{Wallace_1992} and several perceptual hashing algorithms like pHash~\cite{Coskun_Sankur_2004} and Facebook's PDQ~\cite{Davis_Rosen_2019}.

\textbf{Attack intuition.} Our attack exploits the linearity and orthogonality of DCT. By using the Euclidean distance to measure both the input perturbation and the distance between original and modified hashes, we show that minimal perturbations can be found as linear combinations of eigenvectors derived from the linear transform. We further show that the minimal perturbation needed to exceed the threshold $T$ is equal to $T$ exactly.

\textbf{Overview of DCT-based perceptual hashing.} In what follows, we assume that the DCT operates on an image $X$ of size $k \times k$. DCT-based perceptual hashing algorithms indeed typically work by applying a set of transformations to the original image, such as converting to grayscale, resizing, or blurring resulting in a smaller $k \times k$ image. The 2-dimensional DCT (see below) is applied to the $k \times k$ image, resulting in an output of the same size. Next, the dimensionality of the output is reduced by keeping a submatrix of pixels and discarding the rest. For the pHash algorithm, $k=32$ and the $8 \times 8$ submatrix of pixels from the intersection of rows and columns 2 to 9 is preserved, while for PDQ $k=64$ and the $16 \times 16$ submatrix of pixels from the intersection of rows and columns 2 to 17 is preserved. 

\textbf{DCT transform.} The DCT transform for an image $k \times k$ is defined as follows:

$$
h_{DCT}: X \longrightarrow M X M^T
$$
\begin{equation}
 M \in \mathbb{R}^{k \times k},  M_{ij}=\sqrt{\frac{2}{k}} \Lambda_i \cos\Big[\frac{\pi}{k}\Big(j + \frac{1}{2}\Big) i\Big], 1 \leq i,j \leq k
\end{equation}

\begin{equation}
    \Lambda_i=
    \begin{cases}
        \frac{1}{\sqrt{2}}, & \text{if}\ i=0 \\
        1, & \text{otherwise}
    \end{cases}
\end{equation}

$M$ is orthogonal, i.e.,  $M M^T = I_k$. 

The submatrix extraction step applied in perceptual hashing can be written as $h'_{DCT}: X \rightarrow (M X M^T)_{a:b, a:b}$, containing pixels from rows $a$ to $b$ and columns $a$ to $b$, where $1 \leq a \leq b \leq k$ are parameters chosen when developing a perceptual hashing algorithm. $a$ is typically small to extract lower frequencies while $b$ controls the hash size. We rewrite the mapping as $h'_{DCT} : X \rightarrow M' X M'^T$, with $M' = M_{a:b, 1:k} \in \mathbb{R}^{(b-a+1) \times k}$, i.e., $M'$ is the matrix obtained by extracting rows $a$ to $b$ from $M$. Similarly to above, $M' M'^T = I_{b-a+1}$. For more conciseness we set from now on  $c=b-a+1$.

\textbf{Linearity of DCT.} It is straightforward to show that $h'_{DCT}$ is a linear transformation mapping a $k \times k$-dimensional input to a $c \times c$-dimensional output. Simplifying the notation, $h'_{DCT}$ can be rewritten as the linear transform of a vector (the flattened input matrix $X$):

$$ h'_{DCT} : 
    \begin{cases}
        \mathcal{I} \longrightarrow \mathbb{R}^{c^2}\\
        X \longrightarrow A X
    \end{cases}
$$

\begin{gather}
    A \in \mathbb{R}^{c^2 \times k^2}\notag\\
    A_{c_1 \times c + c_2, k_1 \times k + k_2} = M'_{1 + c_1, 1 + k_1} \times M'_{1 + c_2, 1 + k_2}\\
    \text{for } 0 \leq c_1, c_2 \leq c-1, 0 \leq k_1, k_2 \leq k-1 \notag
\end{gather}

An important property of $A$, which we exploit later, is that $AA^T = I_{c^2}$ (see Appendix). This is due to the orthonormality of the rows of $M'$. 

\textbf{Perturbation bound for a given threshold.} We prove that when $f(\delta)\geq T$ then necessarily $||\delta||_2 \geq T$. Using $d$ as the Euclidean distance, we can write $ f_{DCT}(\delta) = ||h_{DCT}(X) - h_{DCT}(X+\delta)||_2 = ||A\delta||_2 = \sqrt{\delta^T A^T A \delta}$. Since $A^T A$ is symmetric, it is diagonalizable in a real orthonormal basis of eigenvectors $(\delta_i)_{1 \leq i \leq k^2}$ by the spectral theorem. The eigenvalues are non-negative because $A^T A$ is positive semidefinite. Since $A A^T = I_{c^2}$, it follows that the eigenvalues can only be 1 or 0. We prove in the Appendix that the multiplicities of eigenvalues are $c^2$ for 1 and $k^2-c^2$ for 0. Without loss of generality, we reorder the eigenvectors by decreasing eigenvalue. Finally, if $\delta = \sum\limits_{i=1}^{k^2} \alpha_i \delta_i$, we can write $f(\delta) = \sqrt{\sum\limits_{i=1}^{c^2}\alpha_i^2} \leq  \sqrt{\sum\limits_{i=1}^{k^2}\alpha_i^2}$. Equality holds if and only if the perturbation $\delta$ belongs to the vector space spanned by eigenvectors $(\delta_i)_{1 \leq i \leq c^2}$.

We provide two practical ways to obtain perturbations $\delta$ such that $f({\delta})\geq T$.

\textbf{Attack as an optimization program.} We frame the attack as the following non-convex optimization program:
\begin{equation}
\label{equation:white-box-optimization}
    \begin{aligned}
        \textrm{Maximize}: &||A\delta||_2^2\\
        \textrm{s. t.}: & ||\delta||_2^2 \leq T^2\\
        &\delta \in \mathcal{I} - X\\
    \end{aligned}
\end{equation}
The first constraint requires that the $\mathbb{L}_2$ norm of the input perturbation is no larger than $T$. The second constraint is linear and requires that the perturbed image is valid. We use the Disciplined Convex Concave Programming toolbox~\cite{shen2016disciplined} to find a solution.

\textbf{Attack using sampling.} This attack samples valid perturbations of norm $T$ \textit{uniformly at random}, using rejection sampling. Alg. \ref{algorithm:white-box-rej-samp} describes our procedure to sample perturbations of norm $T$ along eigenvectors that are not canceled by the linear mapping $(\delta_i)_{1 \leq i \leq c^2}$ 
until a valid perturbation is found meaning that $\delta \in \mathcal{I} - X$. Given a threshold $T$, we sample uniformly at random a vector $\alpha$ on the $\mathbb{L}_2$-sphere of radius $T$ in $c$ dimensions (lines 5-6). We set the perturbation to be $\delta = \sum\limits_{i=1}^{c} \alpha_i \delta_i$ (line 7). Due to the orthonormality of the eigenvector basis, indeed $||\delta||_2^2 = \sum\limits_{i=1}^{c} \alpha_i^2 = T^2$.

\begin{algorithm}[h]
\caption{White-box attack using rejection sampling}
\label{algorithm:white-box-rej-samp}
    \begin{algorithmic}[1]
    \Inputs{
    $T$: Threshold to be attacked.\\
    $(\delta_{i})_{1 \leq i \leq c}$: Orthonormal eigenvectors of $A^T A$ for eigenvalue 1. 
    }

    \Output{
    $\delta$: Minimal perturbation (of $\mathbb{L}_2$ norm $T$) such that $f_{DCT}(\delta)=T$.
    }
    \While{$\delta \notin \mathcal{I} - X$}
        \State{$\alpha \sim N(0, \mathcal{I}_{c})$}
        \State{$\alpha \gets T\frac{\alpha}{||\alpha||_2}$}
        \State{$\delta \gets \sum\limits_{i=1}^{c} \alpha_i \delta_i$}
    \EndWhile
    \end{algorithmic}
\end{algorithm}

\section{Experimental setup}\label{section:experimental_setup}
In this section, we describe the perceptual hashing algorithms used to instantiate our attack and how we implemented it.

\begin{figure*}[htbp]
\centering
\includegraphics[scale=0.42]{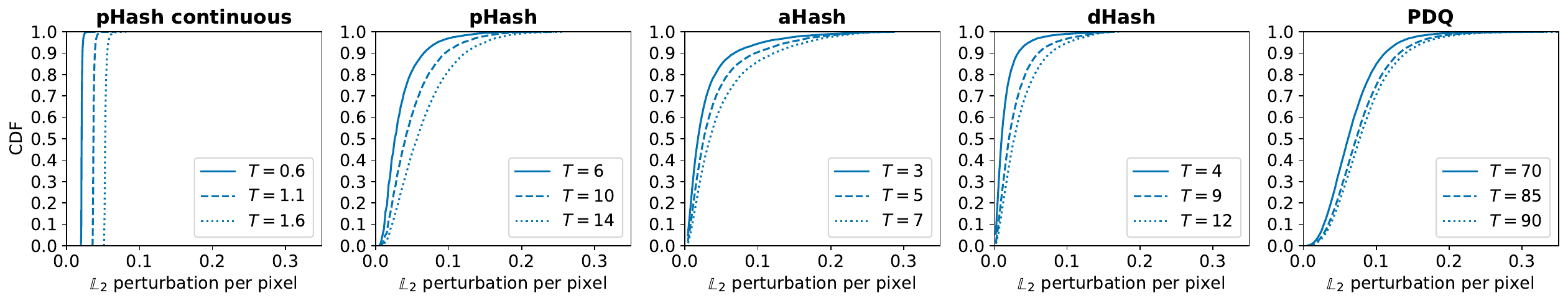}
\caption{Cumulative distribution function (CDF) for the $\mathbb{L}_2$ perturbations per pixel for different algorithms and thresholds and all successfully attacked images over 10 experiments. A lower perturbation indicates higher visual similarity between the modified and original images. The perturbation increases slowly with the threshold, but remains small in all cases.}
\label{fig:l2-per-pixel}  
\end{figure*}

\subsection{Perceptual hashing algorithms}
The perceptual hashing algorithms evaluated in this paper are pHash~\cite{Coskun_Sankur_2004,Zauner_2010}, aHash, dHash~\footnote{https://hackerfactor.com/blog/index.php\%3F/archives/432-Looks-Like-It.html}, and Facebook's PDQ~\cite{facebookpdq}. They are commonly used algorithms for image deduplication or retrieval. 

pHash, aHash and dHash outputs a 64-bit hash while PDQ outputs a 256-bit hash for any input image. The Hamming distance is used to compare the outputs of the hashing algorithms. Each algorithm applies a sequence of transformations such as grayscaling and resizing followed by image feature extraction (e.g. using DCT), and finally a bit discretization step. To experiment with an algorithm with real-valued outputs (of size 64), we remove the discretization step of the pHash algorithm to obtain another algorithm, which we call pHash continuous. The Euclidean distance is used to compare the outputs of pHash continuous.

pHash and pHash continuous apply grayscaling, box blurring and resizing resulting in a $32\times32$ image. They then apply the DCT as explained in Sec. \ref{attack-methodology-white-box}, with parameters $a=1, b=8, k=32$ to get a 64-sized vector. We use this real-valued vector as the output of pHash continuous. pHash assigns bits larger than the median to 1 and 0 otherwise to obtain the final hash.

dHash and aHash apply grayscaling followed by resizing to get a $9\times8$ and $8\times8$ image, respectively. aHash outputs a 64-bit hash by flattening the image, and setting the pixels above mean value to 1 and 0 otherwise. dHash computes a difference between consecutive columns of the image, flattens the computed difference, sets the values greater than 0 to 1, and 0 otherwise to report a 64-bit hash.

PDQ is inspired by pHash and outputs a 256-bit hash. We will not attempt to describe the algorithm in detail, but to provide an overview. PDQ applies grayscaling to the image followed by a series of image transformations to get a $64\times64$ image. It then applies DCT as explained in Sec.~\ref{attack-methodology-white-box} with $a=1,b=16,k=64$, followed by bit quantization by setting bits larger than the median to 1, and 0 otherwise to finally obtain a 256-bit hash.

\subsection{Attack parameters}\label{subsection:implementation-details}
\textbf{Black-box attack.} We use the following parameters to run our black-box attack. All the images are converted to grayscale and resized to $n'=64 \times 64=4,096$ dimensions. We use the $\mathbb{L}_2$ norm to quantify visual similarity and $\epsilon_0 = \eta_{\epsilon}=1/255$. The plateauing behavior of the objective function, i.e., when it stops increasing steadily, is detected based on the last $k=10$ iterations. The number of Gaussian samples used for gradient estimation in each iteration is $d'=800$. We run the attack for a maximum number of iterations $m=10,000$, although in practice only a few hundred iterations are required for most images. We use a momentum $\mu=0$ for all hashing algorithms but pHash continuous for which we use $\mu=0.8$. The values for the scale of Gaussian noise $\sigma$ and learning rate $\eta$ are: $\sigma=0.001, \eta=0.001$ for pHash continuous, $\sigma=0.1, \eta=0.01$ for pHash, $\sigma=0.1, \eta=0.001$ for aHash, $\sigma=0.1, \eta=0.001$ for dHash and $\sigma=0.1, \eta=0.01$ for PDQ. An attacker might want to adjust these parameters for each image, e.g., in order to run the attack more efficiently. For simplicity, we here use the same parameters for all the images. We chose the values of these parameters for each perceptual hashing algorithm based on early experiments on a handful of images from the Stanford Dogs dataset~\cite{khosla2011}.

\textbf{Diversity.} We use the algorithm with modifications for diversity as detailed in~\sref{attack-methodology-black-box}. We set $\epsilon_{0} = 0.25$ and $\eta_{0} = 0.01$. Our results on $\mathbb{L}_2$ perturbations per pixel for different algorithms and different thresholds in Fig.~\ref{fig:l2-per-pixel} show that pHash continuous requires only a small perturbation to generate successful perturbations, while for the other algorithms we need a threshold-specific $\epsilon_{start}$. With these observations, we set $\epsilon_{start}=0$ for pHash continuous for all thresholds. For aHash and dHash we set $\epsilon_{start}=0.03,0.04,0.05$, and for pHash and PDQ we set $\epsilon_{start}=0.06,0.08,0.1$ for three respective thresholds. 

\textbf{Implementation details.} We implement the framework and the perceptual hashing algorithms in Python 3.7~\cite{python} using NumPy~\cite{harris2020array}, OpenCV~\cite{opencv_library}, and SciPy~\cite{2020SciPy-NMeth}. We further use DCCP~\cite{shen2016disciplined} for white-box optimization and Faiss~\cite{JDH17} for efficiently matching hashes against the database.

\section{Results}\label{section:results}

We use the ImageNet dataset from ILSVRC 2012 challenge~\cite{russakovsky2015imagenet} for empirical evaluation of our attacks. For privacy reasons we discard the images containing faces as detected by Yang et al.~\cite{yang2021study}. We remove duplicates from ImageNet using a semi-automated procedure (see Sec.~\ref{appendix:duplicate-analysis} in the Appendix for details), leaving us with a dataset of 1,179,448 images.

\begin{figure*}[htbp]
\centering
\includegraphics[scale=0.43]{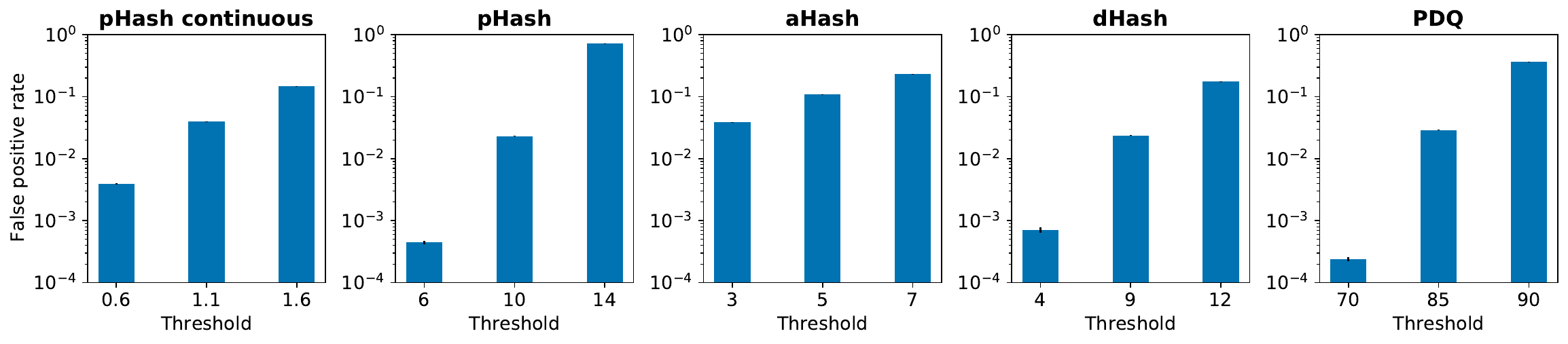}
\caption{False Positive Rate ($FPR$) for different algorithms and thresholds. The database size is $N=100$k, the number of images used to estimate the $FPR$ is $M=500$k and the number of attacked images is $N'=1$k. We show the mean with error bars for the standard deviation (which is very small) over 10 repetitions.}
\label{fig:fpr-n-100k}  
\end{figure*}

For each repetition ($R=10$), we sample two mutually exclusive sets of images, of sizes $N$ and $M$, from ImageNet uniformly without replacement. The images in the first set are used to create the database $\mathcal{D}$ of size $N$ (by default, we use $N=100$k images) and the images in the second set are used as images visually different to those in the database, to estimate the false positive rate (we use $M=500$k images). We select three thresholds for each perceptual hashing algorithm, shown in Table \ref{table:thresholds} (see Discussion for an analysis). We run the black-box attack on $N'=1,000$ randomly sampled images from the database $D$. While the threshold will vary depending on the use case, the lowest threshold would be typically used in practice (see e.g., \cite{facebookpdq} for PDQ). The second and in particular the third threshold would probably generate too many false positives in practice. We include them to test our attack in more challenging scenarios. Given a threshold $T$ and $N'$ images to be attacked, we compute the \textit{attack effectiveness} as the proportion of images that are successfully attacked, i.e., a perturbation is found satisfying $f_X(\delta)>T$.

\begin{table}[htbp]
\centering
\begin{tabular}{|l|l|} 
\hline
 \textbf{Algorithm} & \textbf{Thresholds} \\ 
 \hline
 pHash continuous & 0.6, 1.1, 1.6 \\ 
 pHash  & 6, 10, 14 \\ 
 aHash & 3, 5, 7  \\ 
 dHash & 4, 9, 12 \\ 
 PDQ & 70, 85, 90  \\ 
 \hline
\end{tabular}
\caption{Thresholds selected for the detection system.}
\label{table:thresholds}
\end{table}

In our experiment, our black-box attack manages to successfully attack all images ($N' \times R=10,000$) for all perceptual hashing algorithms but one, reaching an attack effectiveness of 99.9\%. More specifically, we found that one image could not be attacked at the thresholds considered when using aHash for one particular instantiation of our attack . All other images were all successfully attacked for all perceptual hashing algorithms and thresholds considered.

Fig.~\ref{fig:perturbation-images-paper} shows examples of images successfully perturbed using our attack along with the threshold used and resulting $\mathbb{L}_{2}$ perturbation per pixel. The modified image always preserves the visual content of the original image with PDQ seemingly requiring the most visually perceptible modifications. Values of up to $\mathbb{L}_{2, \text{pixel}} \approx 0.10$ resulting in very small changes are typically needed for most images even with the largest thresholds considered. We chose images where the subject is in focus so that the effect of perturbations can be clearly seen and we refer the reader to the Appendix section of the extended version of our paper~\footnote{\url{https://arxiv.org/abs/2106.09820}} for the complete results of our attack against these images for all perceptual hashing algorithms and thresholds.

Fig.~\ref{fig:l2-per-pixel} shows that a small $\mathbb{L}_{2}$ perturbation per pixel is enough to successfully attack most of the images for all hashing algorithms. We obtain similar results using the popular Learned Perceptual Image Patch Similarity (LPIPS) distance~\cite{zhang2018perceptual} (see Fig.~\ref{fig:lpips-distance} in the Appendix) that measures the perceptual distance between images using intermediate layers of a pre-trained image classification model. We can see that the perturbation increases overall with the threshold for all perceptual hashing algorithms. This is expected as a larger perturbation might be needed to push an image at least $T$ away from the original. However, even for the largest thresholds, we observe that an $\mathbb{L}_2$ perturbation per pixel of $0.10$ is enough to successfully attack with imperceptible modification 95.3\% of images for dHash (respectively 100.0\%, 83.5\%, 86.8\% and 73.4\% for pHash continuous, pHash, aHash and PDQ). While results cannot be directly compared across algorithms as threshold are algorithm-specific, we observe interesting difference in the shape of the various algorithm's CDF. pHash continuous seems to requires a similar amount of perturbation for all images (at a given threshold) with, e.g., $\mathbb{L}_2$ for $T=1.6$ only ranging between 0.050 and 0.082. pHash, dHash, and aHash have similar distributions while PDQ seems to always require some changes to the image (resulting in a CDF shifted to the right). We hypothesize the behavior we observe for pHash continuous to be due to two factors: (1) the attack mostly targets the DCT step where it finds an almost optimal $\mathbb{L}_2$ perturbation of $T=1.6$ for an image size of $32 \times 32$, yielding an $\mathbb{L}_2$ perturbation per pixel of $\approx 1.6/32=0.05$ and (2) the other transformations from original to resized images (and from resized to original image to invert the perturbation) roughly preserve the $\mathbb{L}_2$ perturbation per pixel.

Fig. \ref{fig:fpr-n-100k} shows that a detection system would have a very large False Positive Rate ($FPR$), even for the lowest threshold considered. We here empirically estimate $FPR$ as the fraction of images among the $M$ images outside the database flagged by the system. Even for the smallest threshold considered, the false positive rate is larger than 0.02\%. While seemingly small, such values of $FPR$ would, in practice, result in a large number of images being detected by the system and having e.g. to be sent unencrypted to be analyzed. In 2017, 4.5B images were shared daily on Whatsapp alone \cite{whatsappblog2017}. For a prevalence rate of illegal content of $p=10^{-4}$ and a database size of $N=100$k, this would result in >1M false positive images having to be shared unencrypted daily (see Table \ref{table:prevalence}), raising very significant privacy concerns. For the largest thresholds, the false positive rate increases to rates between 14.7\% for pHash continuous to 73.0\% for pHash resulting in hundreds of millions of images being incorrectly flagged and shared daily. We show in the Appendix how this number does not vary much as a function of prevalence rates.

\begin{table}[htbp]
\centering
\begin{tabular}{|cccc|} 
\hline
 \multirow{2}{*}{\textbf{Hash}} & \multicolumn{3}{c|}{\textbf{Rough estimate of the number of}} \\
 & \multicolumn{3}{c|}{\textbf{wrongly detected images daily / }} \\
 & \multicolumn{3}{c|}{\textbf{Threshold $T$}} \\
 \hline
 pHash con- & $\sim18$M & $\sim179$M & $\sim 660$M\\
 tinuous  & $T=0.6$ & $T=1.1$ & $T=1.6$  \\ 
 \hline
  \multirow{2}{*}{pHash}   & $\sim 2$M & $\sim 104$M & $\sim3.3$B\\
   & $T=6$ & $T=10$ & $T=14$  \\ 
 \hline
   \multirow{2}{*}{aHash}  & $\sim 175$M & $\sim 496$M & $\sim1.1$B\\
   & $T=3$ & $T=5$ & $T=7$  \\ 
 \hline
   \multirow{2}{*}{dHash}   & $\sim3$M & $\sim 106$M & $\sim796$M\\
   & $T=4$ & $T=9$ & $T=12$  \\ 
 \hline
  \multirow{2}{*}{PDQ} & $\sim1$M & $\sim130$M & $\sim1.6$B\\
  
  & $T=70$ & $T=85$ & $T=90$  \\ 
 \hline
\end{tabular}
\caption{Rough estimate of the number of images that would be wrongly detected daily on WhatsApp alone for different perceptual hashing algorithms and thresholds. We here consider a database size of $N=100$k, 4.5B images being shared daily \cite{whatsappblog2017}, and a prevalence rate of illegal content of $10^{-4}$.}
\label{table:prevalence}
\end{table}

\begin{figure*}[htbp]
\centering
\includegraphics[scale=0.42]{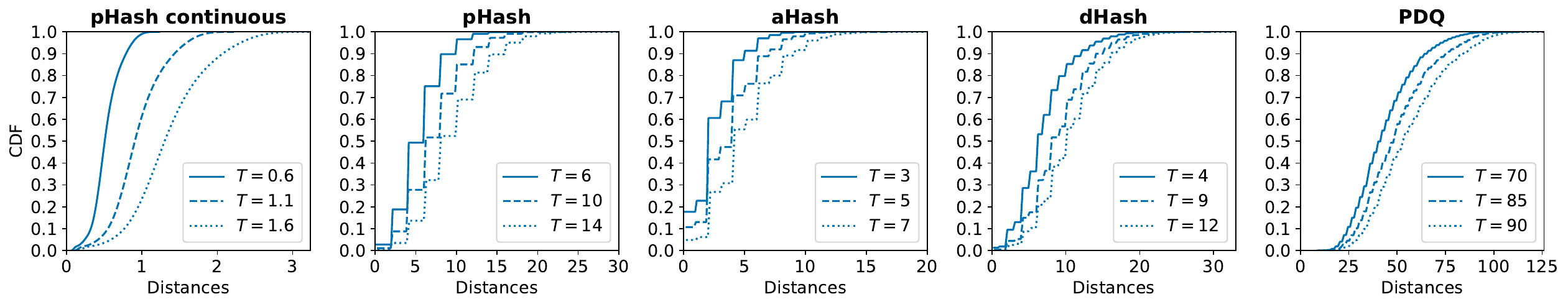}
\caption{Pairwise distances between hashes of the multiple modified images generated for the same image, for different algorithms and thresholds. $D = 50$ modified images are generated for each ($N' = 100$) original image.}
\label{fig:diverse-perturbation-distances}  
\end{figure*}

\begin{figure}[htbp]
\setlength{\belowcaptionskip}{-1\baselineskip}
\centering
\includegraphics[scale=0.41]{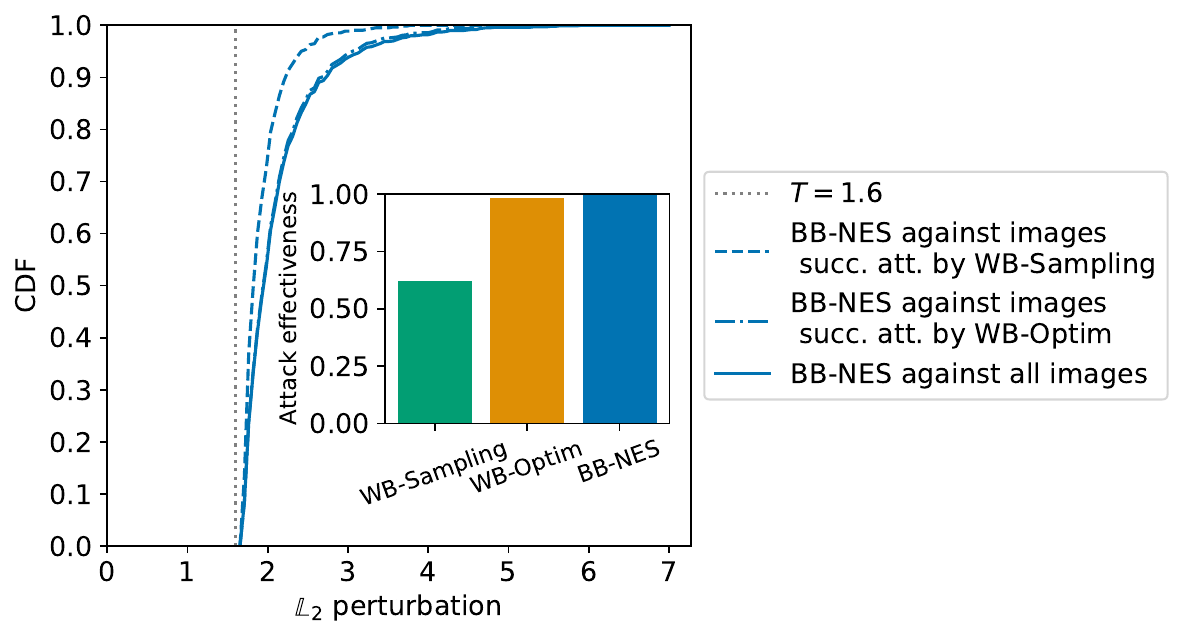}
\caption{Cumulative distribution function (CDF) for the $\mathbb{L}_2$ norm of perturbations using our black-box attack (BB-NES) against the DCT step and a threshold $T=1.6$ (gray dotted line) for $N'=1,000$ images. The inset shows the attack effectiveness for the three attacks.}
\label{fig:optimality-black-box}  
\end{figure}

\textbf{Diversity.} Contrary to the ML case, ensuring that an attack generates diverse enough perturbations every time it is run is essential to make it a practical concern for client-side scanning. Deterministic attacks could indeed easily be thwarted by expanding the database to include the original images and its modifications. To assess the ability of our attack to produce different perturbations, we attack $N''=100$ images sampled uniformly at random without replacement from the database. Each image is attacked $D=50$ times, using the diversity-focused version of the attack and different random initializations (seeds) (see~\sref{attack-methodology-black-box}). Fig. \ref{fig:diverse-perturbation-l2} (see Appendix) shows the $\mathbb{L}_2$ perturbation per pixel is slightly larger than in the default black-box attack. Yet, most perturbations are imperceptible and have $\mathbb{L}_2$ perturbation per pixel $\leq 0.15$ for all thresholds and algorithms.

Fig.~\ref{fig:diverse-perturbation-distances} shows the pairwise distances between hashes of modified images generated from the same image. The distances between modified images are roughly similar to the threshold used for all the perceptual hashing algorithms evaluated. This suggests that the modified images are well distributed around the ball centered on the original image and not clustered in one part of the space. This also strongly suggests that simple mitigation strategies like expanding the database would not be sufficient to counter the attack.

Finally, we assess the capacity of the black-box approach to produce optimal perturbations against the DCT step. We compare the results of the black-box attack to those of the white-box attacks and to our theoretical lower-bound (for a threshold $T$ and the Euclidean distance, the $\mathbb{L}_2$-norm $\geq T$, see Sec. \ref{attack-methodology-white-box}). We attack $N'=1,000$ images randomly sampled from ImageNet after resizing them to $k \times k$ with $k=32$ and use the black-box attack parameters from the DCT-based pHash continuous. For the WB-Sampling approach, we generate up to 1M samples and we consider the attack successful as soon as $\delta \in I - X$ (see Alg. \ref{algorithm:white-box-rej-samp} for details). For the WB-Optim approach, we run the optimization program for a maximum of 1,000 iterations and for up to 10 times, until a solution is found; if none is found, the attack is considered unsuccessful.  Fig.~\ref{fig:optimality-black-box} shows that the black-box attack achieves an input perturbation no larger than 2 times the theoretical limit 96\% of the time and no more than 0.5 above the limit 74\% of the time. Interestingly, in this experiment, while the black-box attack always succeeds in finding a perturbation, the white-box approaches are less successful. WB-Optim and WB-Sampling achieve an attack effectiveness of only 98.4\% and 62.2\% respectively. We believe that WB-Sampling is less successful because the sampled perturbations are more likely to go out of the image bounds for some images, resulting in no successful perturbation. Finally, we note that the $\mathbb{L}_2$ perturbation is smaller for the images successfully attacked by WB-Sampling, perhaps because it easier to devise small perturbations for these images.

\begin{figure*}[htbp]
\centering
\includegraphics[scale=0.43]{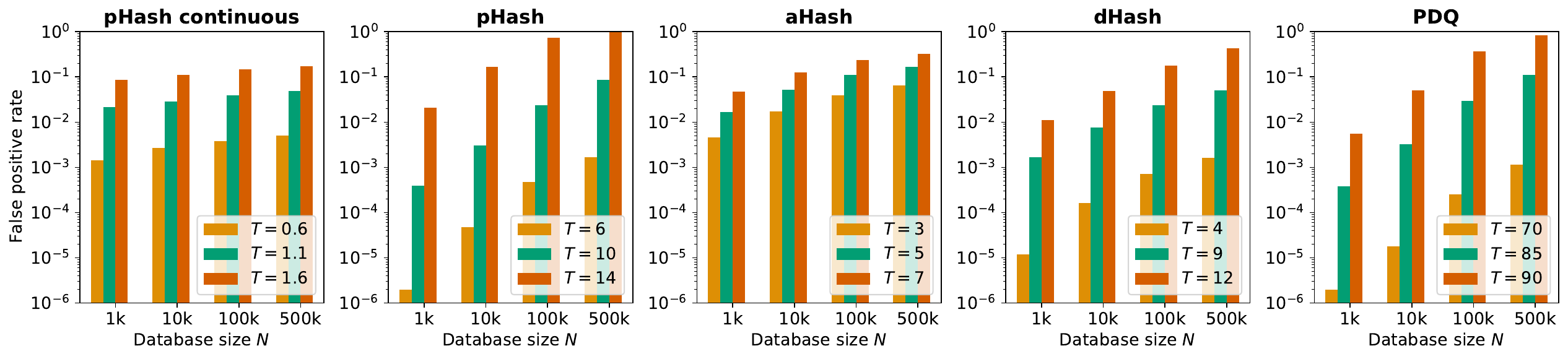}
\caption{False Positive Rate ($FPR$) for each algorithm, threshold, and database size $N$. The number of images visually different from those in the database that are used to estimate the $FPR$ is $M=500$k.
}
\label{fig:fpr-increasing-n}  
\end{figure*}

\section{Discussion}
\label{section:discussion}

Our results and, in particular, the $FPR$s for specific thresholds are estimated using ImageNet. While ImageNet already contains more than 1M images, results might differ in different and, in particular, larger datasets. We randomly sample $P$ images from ImageNet and, for each hashing algorithm, compute the distances between image hashes for up to 1M pairs. We find the distribution of distances between images to be roughly stable when the number of images varies for each algorithm. Similarly, we run the analysis on another dataset, Stanford Dogs~\cite{khosla2011}, and show the distribution of distances to be similar to that of ImageNet and stable. This suggest that our $FPR$s can be safely extrapolated to larger and different datasets. Fig.~\ref{fig:cdf-distances} in the Appendix supports this analysis. 

We have, throughout this work, assumed that the attacker knows the threshold and distance function used by the system to flag images. We believe this assumption to be reasonable as a) pairwise distances are fairly similar across datasets and stable, allowing an attacker to reasonably estimate acceptable thresholds and b) our attack works well, producing images very similar to the original one, even with very high thresholds. We also believe the assumption that the attacker knows the distance function used to be reasonable, as an attacker is likely to be able to infer the distance being used based on the output values, especially with the Euclidean and Hamming distance being the most commonly used.

While we report results on a database size of $N=100$k, larger databases are often used in practice (e.g., the National Center for Missing and Exploited Children in the US reported a database size of 47.2M~\cite{iicsa2020} and the Global Internet Forum for Counter Terrorism reported a database of size 250k~\cite{gifct2020}). Fig. \ref{fig:fpr-increasing-n} shows that the $FPR$ of a system increases with the database size $N$. Even for the lowest threshold, the $FPR$ for $N=500$k reaches 0.51\% for pHash continuous ($T=0.6$), 0.17\% ($T=6$) for pHash, 6.44\% ($T=3$) for aHash, 0.16\% ($T=4$) for dHash and 0.11\% ($T=70$) for PDQ. This strongly suggest that the $FPR$s we report are a lower bound on the $FPR$s and therefore the privacy risk in practice.

\begin{figure*}[htbp]
\centering
\includegraphics[scale=0.43]{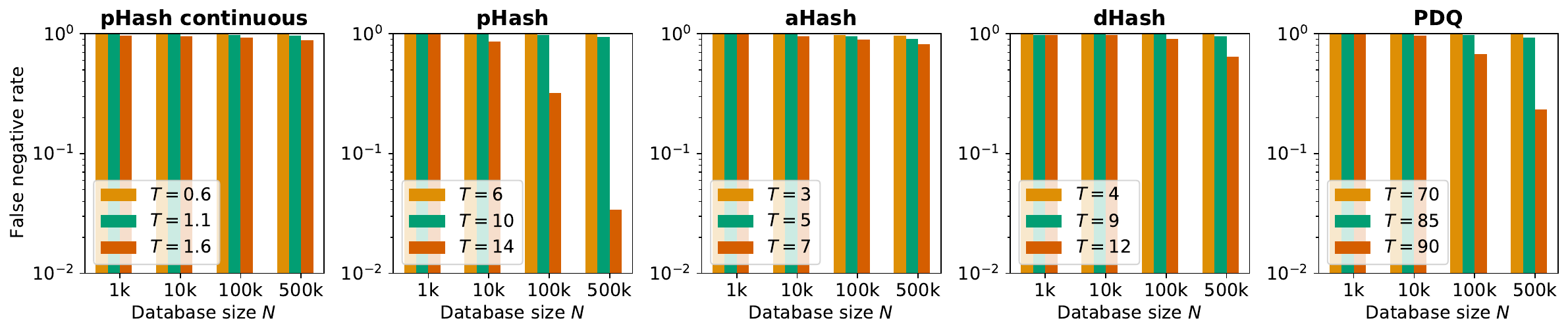}
\caption{False Negative Rate ($FNR$) for each algorithm, threshold, and database size $N$. The number of attacked images used to estimate the $FNR$ is $N'=1,000$.}
\label{fig:fnr-increasing-n}  
\end{figure*}

We have so far considered our attack to be successful when the modified image is at a distance $>T$ from the original image. In practice, however, a modified image might be at a distance $>T$ from the original image but still be at a distance $\leq T$ from another image in the database and be (correctly, even if for the wrong reason) flagged. To evaluate the impact this has on the detection avoidance capability of our attack, we compute the False Negative Rate ($FNR$), the fraction of modified images incorrectly rejected by the system. Fig. \ref{fig:fnr-increasing-n} shows that the $FNR$ stays extremely high for all perceptual hashing algorithms at reasonable thresholds, indicating that very few modified images get flagged for being at a distance $\leq T$ from another image in the database. As expected, $FNR$ decreases (and $FPR$ increases) when databases becomes large and a large threshold is used, leading to a significant fraction of the space being considered ``close'' to at least one illegal image. For instance, for a database size $N=500$k and the highest threshold, 3.4\% of modified images would still evade detection by pHash and 23.4\% by PDQ.

Our black-box attack relies on a heuristic to minimize the perturbation while increasing the distance to the original image above a threshold $T$. Although we produce imperceptible perturbations for most of the attacked images, our results are only an upper estimate of the actual minimum perturbation. Prior work studying the robustness of machine learning classification models to adversarial perturbations showed that finding minimal perturbation is a hard problem~\cite{katz2017reluplex} and even lower perturbations might be achieved in future work.

We focus, in our white-box attack, on DCT-based perceptual hashing algorithms, exploiting the linearity of the DCT transform. Our attack is likely to be extendable to other linear image transforms, e.g. image scaling \cite{xiao2019seeing}, opening the door to future attacks as well as potentially novel theoretical insights.

\textbf{Countermeasures. } We discuss several countermeasures that a detection system could implement to thwart our attack. 

First, the system could use a larger threshold. We showed in Sec.~\ref{section:results} that the larger thresholds are not only ineffective in detecting adversarially modified images but also lead to a significant increase in false positives.

Second, one could use our attacks to generate hashes and add them to the database. We show in Sec.~\ref{section:results} that our attack is able to generate diverse perturbations so as to prevent such countermeasures. 

Third, systems (such as Apple's protocol~\cite{applechildsafety}) could flag users only after the number of matches exceed a predefined threshold $k$, rather than on a single match. Using a simple model, we show that such a measure is not a trivial countermeasure to our attack. 

\begin{figure}[htb]
\setlength\abovecaptionskip{-1\baselineskip}
\centering
\includegraphics[scale=0.43]{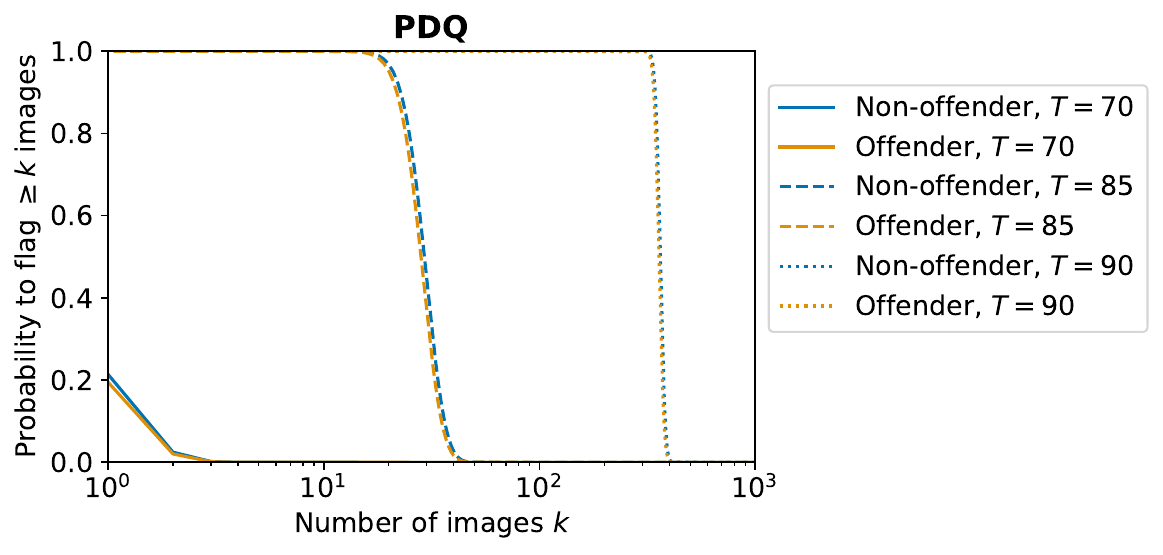}
\caption{Probability that a detection system flags at least $k$ images shared by non-offenders (sharing 0 illegal images) and offenders (sharing 100 illegal images). Both offenders and non-offenders share a total of 1,000 images. 
}
\label{fig:csp-atleast-k-pdq}  
\end{figure}

More specifically, we assume that both offenders and non-offenders send 1000 images to the server. However, while a) non-offenders send 0 illegal images  b) offenders send exactly 100 illegal images (10\%). For each type of user and each threshold, we compute the probability for the system to flag $1 \leq k \leq N$ images using the estimated $FPR$ and $FNR$ for a database size of 100k (see Appendix for the exact formulas). Fig~\ref{fig:csp-atleast-k-pdq} shows that offenders and non-offenders are similarly likely to have at least $k$ images flagged. We obtain similar results for the other perceptual hashing algorithms (see Fig. ~\ref{fig:csp-at-least-k-all} in the Appendix). This can be attributed to the fact that the probability for a non-illegal image to get flagged ($FPR$) is comparable to the probability for an adversarially modified illegal image to be flagged (true positive rate or $TPR = 1-FNR$). These results suggest that flagging a user with at least $k$ matches is not a trivial countermeasure against our attack. We leave a deeper analysis of such a mechanism for future work. 

Fourth, the detection system could apply additional image transformations before computing the perceptual hash of the image. Such modifications would however be part of the black-box and attacked by our algorithm. Simple transformations are thus unlikely to prevent our adversarial detection avoidance attack. Similarly, modifying the perceptual hashing algorithm for instance by using a secret set of DCT coefficients that is different from the set used by pHash or PDQ is unlikely to prevent our attack. These would furthermore make the algorithm less robust as the coefficients extracted by pHash and PDQ represent the most important features of the image. 

Fifth, increasing the length of the hashes is likely to help better distinguish between different images and reduce the false positive rate. This might allow the system to increase what are realistic thresholds and make our attack harder. We here use hashes of size 64 (and 256 for PDQ \cite{facebookpdq}, see Sec. \ref{subsection:implementation-details}). Most perceptual hashing algorithms can be adapted to provide hashes of different sizes. For instance, pHash, aHash and dHash can all output hashes of size 256 \footnote{https://github.com/thorn-oss/perception}. While further analysis is required, we note here that a) our attack works well even for large thresholds and b) that longer hashes encode more information about the image. This raises strong privacy concerns e.g. reversal attacks~\cite{locascio2018} and, in the extreme, defeats the purpose of perceptual hashing.

\textbf{Implications.} Our research shows that current perceptual hashing (PH) algorithms are not robust to black-box detection avoidance attacks and that no straightforward mitigations strategies exist. We believe PH algorithms, which are designed to produce hashes that change gradually as the image changes, might be inherently vulnerable to attacks. Our results, combined with the concerns PH-CSS and in particular cryptographic-enhanced PH-CSS such as Apple’s~\cite{applechildsafety} raise on the ``illegal'' content being searched for, led us to believe that even the best PH-CSS proposals today are not ready for deployment.

\section{Related work}
\label{section:related_work}

\textbf{Robustness of perceptual hashing algorithms.} Previous work have shown perceptual hashing algorithms to be robust to small standard image transformations like resizing, recoloring, watermarking, cropping, and blurring~\cite{zauner2011rihamark, drmic2017evaluating}.  They evaluate the performances of perceptual hashing algorithms against modifications in duplicate image detection setup, i.e. matching against a single image. They however focus on standard image transformations and do not consider adversaries with more sophisticated tools at their disposal. 

Dolhansky and Ferrer~\cite{dolhansky2020adversarial} studied collision (false positive) attacks in perceptual hashing algorithms under white-box assumptions. More specifically, their algorithm modifies an image such that its hash is same as the hash of a given target image. They conclude that perceptual hashing algorithms should remain secret. In this work, we instead focus on the client-side use case proposed recently by researchers and policy makers~\cite{eu2020doc, whatsapp2019fact} where the attacker has a black-box access to the algorithm. Furthermore, our attack is a false negative attack, aiming at finding minimal diverse perturbations such that distance between the hashes of modified and input image is greater than a given threshold.

\textbf{Adversarial attacks against ML models.} Adversarial attacks have been widely studied against ML models~\cite{ ren2020adversarial, sharif2016accessorize} under both white-box~\cite{szegedy2013intriguing, Goodfellow_Shlens_Szegedy_2015, kurakin2016adversarial,carlini2017towards} and black-box assumptions~\cite{papernot2017practical, liu2016delving}. Image classification models have been particularly found to be vulnerable to adversarial attacks~\cite{akhtar2018threat} leading the image to be misclassified by the ML model. Our attack leverages previous work in adversarial ML including Natural Evolutionary Strategies (NES) (see Sec. \ref{attack-methodology-black-box}).

Adversarial attacks on ML models in a black-box setting assume that the attacker has an access to query the model and get the output. A commonly observed approach is to train a substitute model to emulate the target model, and then attack the surrogate model~\cite{papernot2017practical, shi2019curls}. This approach implicitly assumes that the surrogate model has enough parameters to provide a good approximation of the target model. This would imply that not all substitute models would work for all the target models and hence the attack might not always work. 

Another approach is to iteratively perturb the image and query the model to update the perturbation in each iteration~\cite{guo2019simple}. These methods often use gradient estimation to estimate the perturbation update in each iteration. Natural evolutionary strategies (NES) for gradient estimation is one of the state-of-the-art methods in black-box adversarial ML~\cite{cheng2018query, Ilyas_Engstrom_Athalye_Lin_2018, meunier2019yet}. Our black-box attack also uses NES-based strategy for gradient estimation and perturbation update. Lastly, we note that generating multiple perturbations for the same image for diversity in the output space is not a requirement in the adversarial ML setup, while it is an important requirement for adversarial attacks against PH-CSS.

\section{Conclusion}
\label{section:conclusion}

In this paper, we introduced the first framework to evaluate the robustness of perceptual hashing-based client-side scanning against adversarial attacks. We proposed a general black-box attack and showed that $>99.9\%$ of images can be successfully modified while preserving the image content. We also show our attack to generate diverse perturbations preventing straightforward mitigation strategies such as expanding the database with modified images. We finally propose two white-box attacks, providing a theoretical basis for attacks. 

Taken together, our results shed strong doubt on the robustness to adversarial black-box attacks of perceptual hashing-based client-side scanning as currently proposed. The detection thresholds necessary to make the attack harder are likely to be very large, probably requiring more than one billion images to be wrongly flagged daily, raising strong privacy concerns.

\section*{Acknowledgements} 

We thank the Computational Privacy Group and, in particular, Florimond Houssiau and Ali Farzanehfar for their helpful feedback and comments on the manuscript. We also would like to thank the reviewers, chairs, and our shepherd for their useful feedback and help improving the paper. Finally, we thank the artifact reviewers for pointing out that ImageNet has duplicates, an issue which we address in the current version of the paper by removing the duplicates.

\bibliographystyle{plain}
\bibliography{references}

\appendix

\section{Proofs for the DCT analysis}
\begin{proposition} $A A^T = I_{c^2}$, where A is defined as follows:

\begin{gather}
    A \in \mathbb{R}^{c^2 \times k^2}\notag\\
    A_{c_1 \times c + c_2, k_1 \times k + k_2} = M'_{1 + c_1, 1 + k_1} \times M'_{1 + c_2, 1 + k_2} \notag\\
    \text{for } 0 \leq c_1, c_2 \leq c-1, 0 \leq k_1, k_2 \leq k-1 \notag
\end{gather}
\end{proposition}

\begin{proof}
Let $c = c_1 \times c + c_2$ and $c' = c_1' \times c + c_2'$. We develop $(A A^T)_{c, c'}$ as follows:

\begin{align*}
    (A A^T)_{c, c'} &= \sum\limits_{k_1 = 0}^{k-1} \sum\limits_{k_2 = 0}^{k-1} A_{c_1 \times c + c_2, k_1 \times k + k_2} \times A_{c'_1 \times c + c'_2, k_1 \times k + k_2} \\
    &= \sum\limits_{k_1 = 0}^{k-1} \sum\limits_{k_2 = 0}^{k-1} M'_{1+c_1, 1+k_1} \times M'_{1+c_2, 1+k_2} \times M'_{1+c'_1, 1+k_1} \times \\
    & \times M'_{1+c'_2, 1+k_2} \\
    &= ( \sum\limits_{k_1 = 0}^{k-1} M'_{1+c_1, 1+k_1} \times M'_{1+c'_1, 1+k_1}) \times \\
    & \times (\sum\limits_{k_2 = 0}^{k-1} M'_{1+c_2, 1+k_2} \times M'_{1+c'_2, 1+k_2}) \\
    &= (M' M'^T)_{1+c_1, 1+c_2} \times (M' M'^T)_{1+c'_1, 1+c'_2} 
\end{align*}

Because $M' M'^T = I_{c}$, it follows that (A $A^T)_{c, c'} = 1$ if $c = c'$ and 0 otherwise. Indeed, $c = c'$ if and only if $c_1 = c_1'$ and $c_2 = c_2'$, meaning that the terms being multiplied in the last equality can both be non-zero only when $c=c'$, in which case they are both equal to 1.
\end{proof}

\begin{proposition}
The eigenvalues of $A^T A$ are 1 with multiplicity $c^2$ and 0 with multiplicity $k^2 - c^2$. 
\end{proposition}
\begin{proof}
Let $\lambda$ be an eigenvalue of $A^T A$ and $x \neq 0$ such that $A^T A x = \lambda x$. We multiply by $A$ to the left and obtain $\lambda A x = A (A^T A x) = (A A^T) A x = A x$. We distinguish two cases: (1) $A x = 0$, which implies that $\lambda = 0$ and (2) $A x \neq 0$, which implies that $\lambda = 1$.

Therefore the eigenvalues of $A A^T$ can only be 0 or 1.

We denote by $m(\lambda)$ the multiplicity of eigenvalue $\lambda$. 

It follows from the above analysis that $dim(Ker(A)) \leq m(0)$ and that $rank(A) \leq m(1)$. By summation we obtain $k^2 \leq m(0) + m(1) = k^2$, therefore $m(1) = rank(A)$. 

We show that the rank of $A$ is $c^2$. Indeed $rank(A) \leq min(c^2, k^2) = c^2$. If it were the case that $rank(A) < c^2$, then there would be $x \in \mathbb{R}^{c^2}$ such that $x^T A = 0$ but $x \neq 0$. Since $A A^T = I_{c^2}$, by multiplying with $A^T$ on the right we obtain $x^T = 0$, a contradiction.

We thus conclude that $m(1) = c^2$ and $m(0) = k^2 - c^2$.
\end{proof}

\section{Duplicate analysis}\label{appendix:duplicate-analysis}

This paper's empirical results use a subset of the ImageNet dataset (train, validation, and test combined) where all the images predicted by~\cite{yang2021study} to contain human faces were removed. We also remove duplicates as described below. The cleaned dataset consists of 1,179,448 images. 

We removed duplicates from ImageNet following feedback from the USENIX '22 Artifact Evaluation reviewers. 
While the dataset website claims duplicates were removed, a manual inspection revealed that this was not the case. 
To accurately estimate the $FPR$ of client-side scanning systems at low thresholds, we now remove duplicates. 
We consequently also only report empirical results using the three largest, the hardest to attack, thresholds from the USENIX '22 version. 
The $FPR$s obtained for the lowest threshold were indeed now close to or equal to 0 making it, in our opinion, not a reasonable threshold to use in practice when detecting illegal content. 
All our claims remain the same.

To detect exact and near-duplicates, we use a semi-automated procedure based on standard perceptual hashing algorithms: we flag potential duplicates using perceptual hashing algorithms, manually label these as either true or false duplicates, and finally remove the true duplicates. In total, 8526 images duplicates were removed from the dataset.

\begin{figure}[htbp]
\centering
\includegraphics[scale=0.5]{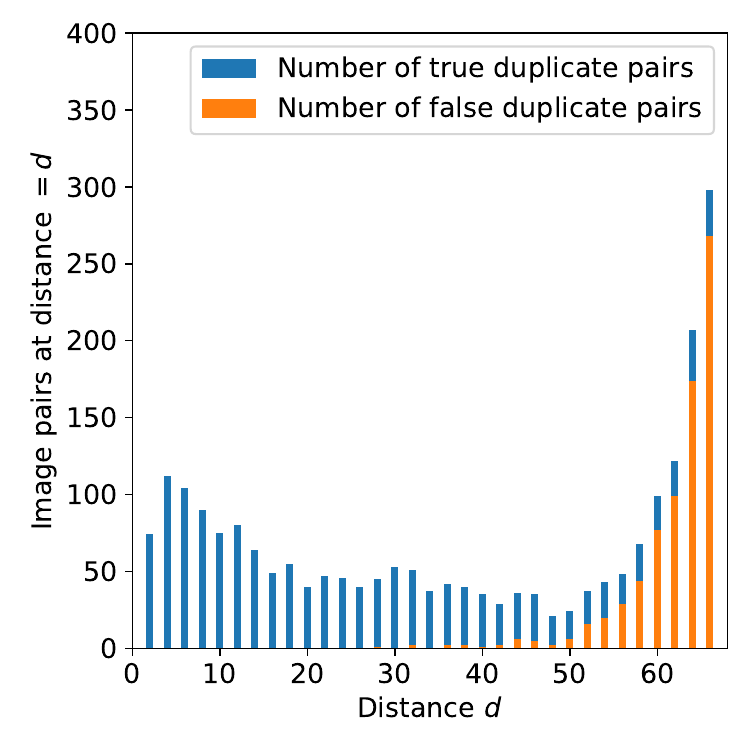}
\caption{Number of potential duplicates detected for each distance $d$ by the PDQ algorithm.}
\label{fig:duplicate-statistics}  
\end{figure}

\textbf{Detecting duplicates.} We start by retrieving for each image their matches at distance $d$ calculated using PDQ. To ensure that we do not overlook duplicates that could potentially result in overestimated false positive rates, we iterated over all the distances $d$ between 0 and 70, the lowest threshold attacked for PDQ. We traverse ImageNet in decreasing image size order, to ensure that potential duplicates are lower-quality images. We manually label each duplicate as either ``true'' or ``false'' and only remove the true duplicates. We acknowledge that the notion of ``true'' duplicate can be subjective, especially for near-duplicates. Similarly to~\cite{douze20212021}, we considered two images where one is an edited copy of the other to be duplicates. Another edge case consists of pictures clearly taken within seconds of one another that are very highly similar; we also considered them to be duplicates. All edge cases were labeled by the three authors.

Figure~\ref{fig:duplicate-statistics} shows the number of potential duplicates detected for each distance d by PDQ (for clarity, we do not display the value for $d=0$ for which 7136 are detected). As it is unfeasible to inspect all of the images at $d = 0$, we inspected a sample of 100 images with duplicates and found all of them to be true (exact) duplicates. For d ranging from 2 to 54, we found a total of 1404 matches which we manually inspected. We found most of them (1339) to be true duplicates, as shown on Figure \ref{fig:duplicate-statistics}. For $d$ ranging from 56 to 66, we found a total of 842 potential duplicate pairs. We manually inspect them and found 151 of them to be true duplicates which we remove. The larger thresholds ($d=68$ and $d=70$) contained more than 400 duplicate pairs. A manual inspection of a random subset revealed that a large majority of them (93\%, resp. 96\%) are false duplicates. Considering the large number of images that would have to be manually inspected for a likely minimal impact we did not remove further images. As the number of false duplicates is already very large, we believe that the small remaining fraction of duplicates is unlikely to significantly impact the $FPR$ of the client side scanning system. 

\textbf{Other hashing algorithms.} After the above procedure and to avoid potential biases, we also check for duplicates using the other hashing algorithms. For each algorithm, we retrieve matches at distance d ranging from 0 to the corresponding lowest threshold attacked in this paper. We do so, in order, for pHash, pHash continuous, aHash, and dHash.
\begin{itemize}
    \item For pHash, 15 ($d=0$) and 69 ($d=2$) images are flagged a potential duplicates. Out of these, 24 are true duplicates which we remove. For $d=4$ and $d=6$, as more than 400 image pairs would have to be inspected (namely 489 and 2950), we sample 100 random images with duplicates and find 99\% and 97\% of them to be false duplicates. As before we stopped here.
    \item For pHash continuous, we retrieve all images with potential duplicates within distance intervals [0, 0.1), [0.1, 0.2), up to [0.5, 0.6). We retrieved 4 matches within the first interval and none of which was a true duplicate. We retrieved 519 matches within the second interval and, upon manual inspection of 100 pairs, we found no true duplicate. For the remaining intervals, we retrieved 5k, 20k, 42k, and 79k. We manually inspected 100 pairs in each of these and found no true duplicate. In fact, in many cases, a large number of visually very different matches were retrieved, indicating that the thresholds ranges were already very high for duplicate detection. 
    \item For aHash, we found a very high number of matches, e.g., 59k for $d=0$ and 140k for $d=3$, the lowest threshold attacked in this paper. We inspected 500 pairs for $d=0$ and found only one true duplicate. We stop here.
    \item For dHash, we find 62 matches for $d=0$ and 191 $d=1$. We manually inspect them and find that only one of these is a true duplicate which we remove. We then find 677 matches for $d=2$, 2125 matches for $d=3$, and 6184 matches for $d=4$, the lowest threshold attacked. We inspected 100 pairs for each threshold and found no true duplicate. 

\end{itemize}

\section{Visual similarity}

\begin{figure*}[htbp]
\centering
\subfigure{
    \includegraphics[scale=0.42]{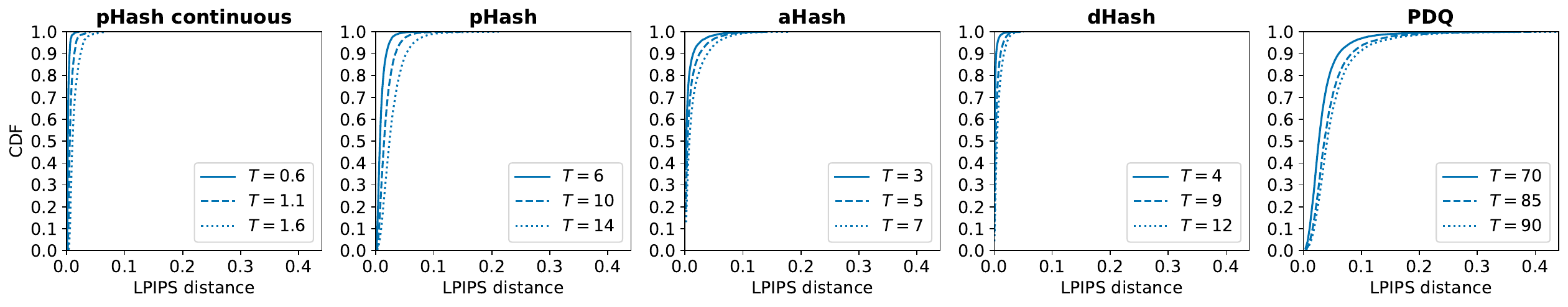}
}
\caption{Cumulative distribution function (CDF) for the LPIPS distance between the original and the perturbed image for different algorithms and thresholds $T$, and all successfully attacked images over 10 experiments. A smaller distance indicates higher visual similarity between the modified and original images. The distances increase slowly with the threshold but remain small in all cases.}
\label{fig:lpips-distance}  
\end{figure*}

Fig. \ref{fig:lpips-distance} shows the perceptual similarity between the original and modified images as quantified by the Learned Perceptual Image Patch Similarity distance~\cite{zhang2018perceptual} evaluated using AlexNet~\cite{alexnet}. 

\arxiv{
    \begin{figure*}[htbp]
    \centering
    \subfigure{
        \includegraphics[scale=0.4]{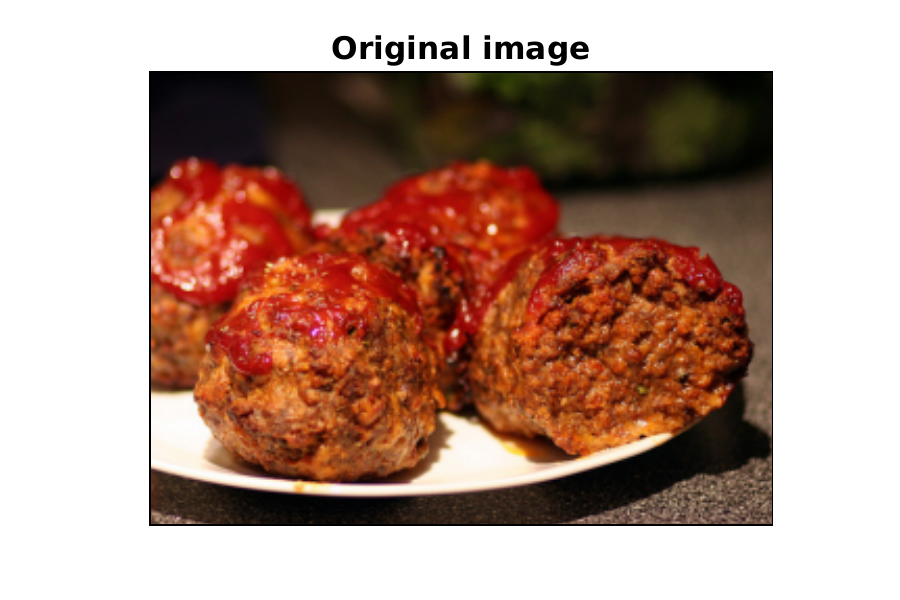}
    }
    \subfigure{
        \includegraphics[scale=0.42]{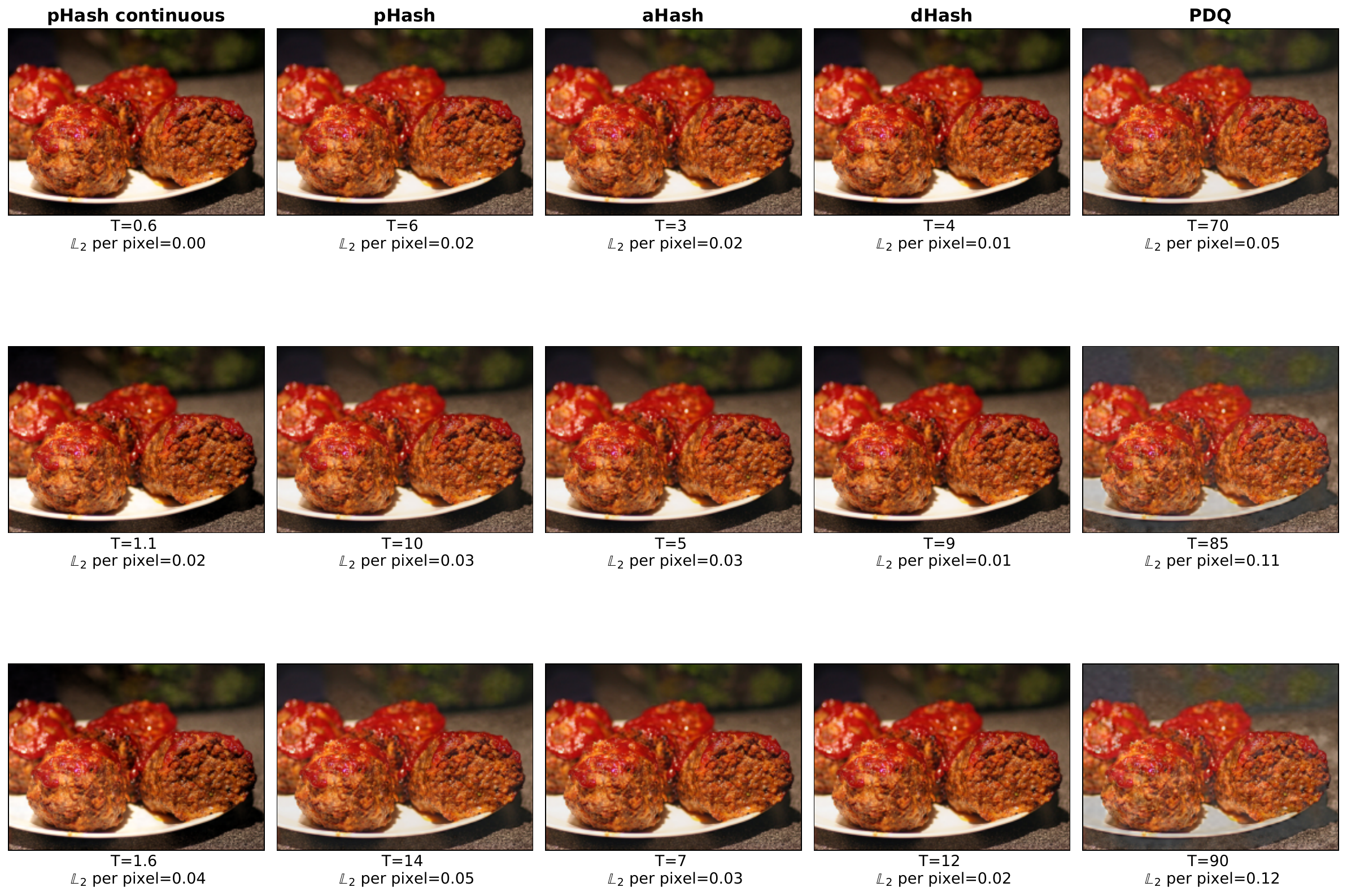}
    }
    \caption{Original image with modifications obtained using our black-box attack for different perceptual hashing algorithms and thresholds. We show the $\mathbb{L}_2$ perturbation per pixel. All the modified images are qualitatively very similar to the original image.}
    \label{fig:modified-images-1}  
    \end{figure*}
    
    \begin{figure*}[htbp]
    \centering
    \subfigure{
        \includegraphics[scale=0.4]{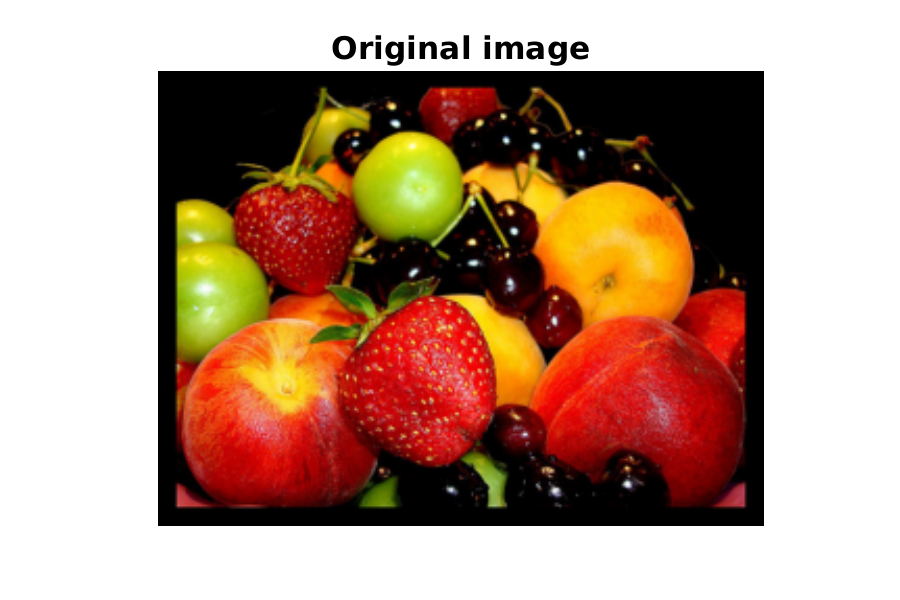}
    }
    \subfigure{
        \includegraphics[scale=0.42]{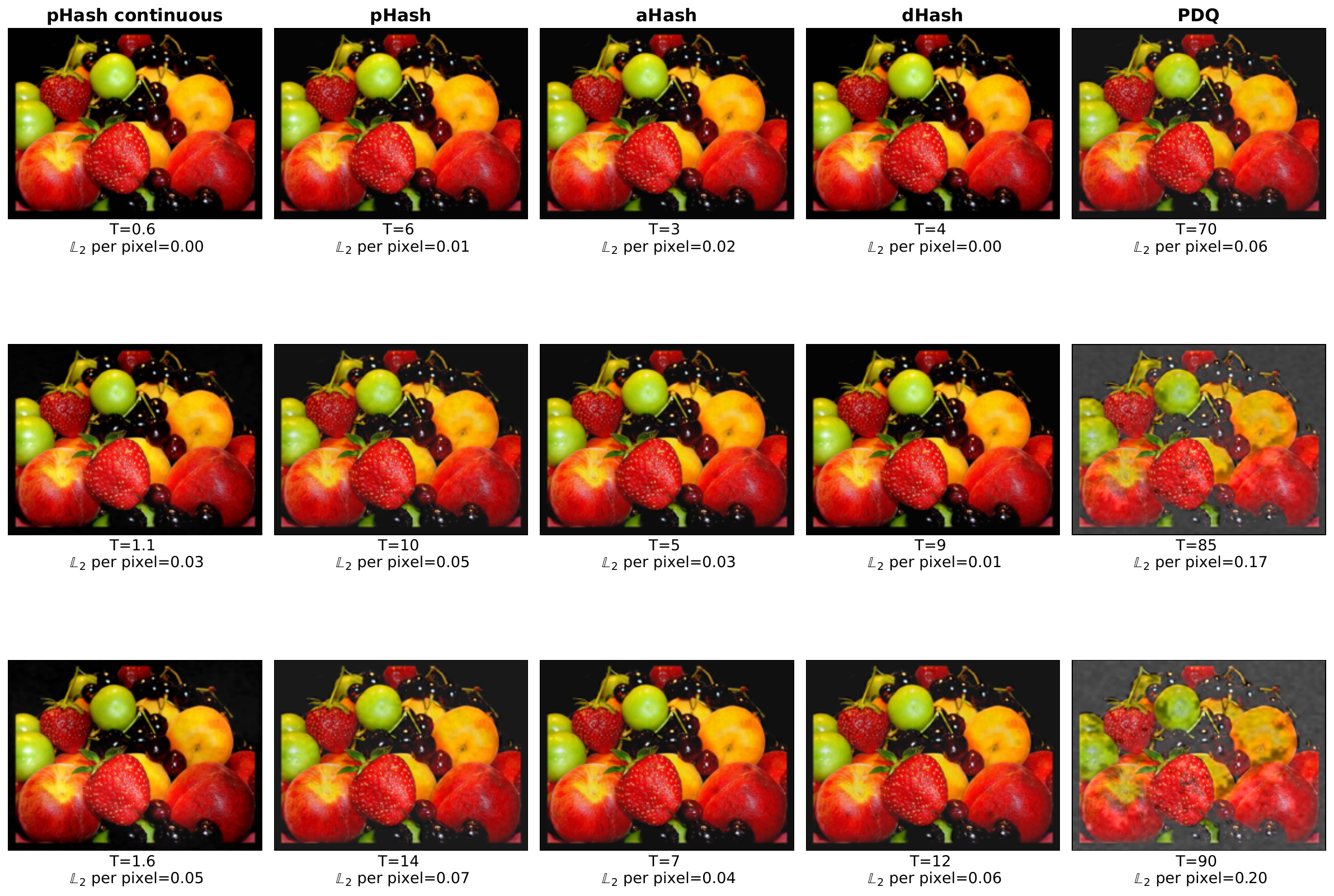}
    }
    \caption{Original image with modifications obtained using our black-box attack for different perceptual hashing algorithms and thresholds. We show the $\mathbb{L}_2$ perturbation per pixel. All the modified images are qualitatively very similar to the original image.}
    \label{fig:modified-images-2}  
    \end{figure*}
    
    \begin{figure*}[htbp]
    \centering
    \subfigure{
        \includegraphics[scale=0.4]{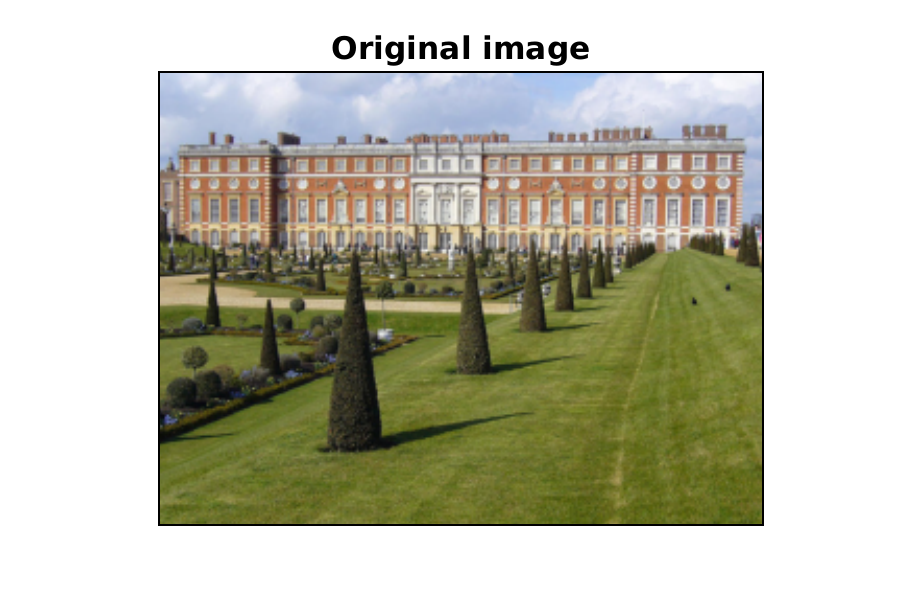}
    }
    \subfigure{
        \includegraphics[scale=0.42]{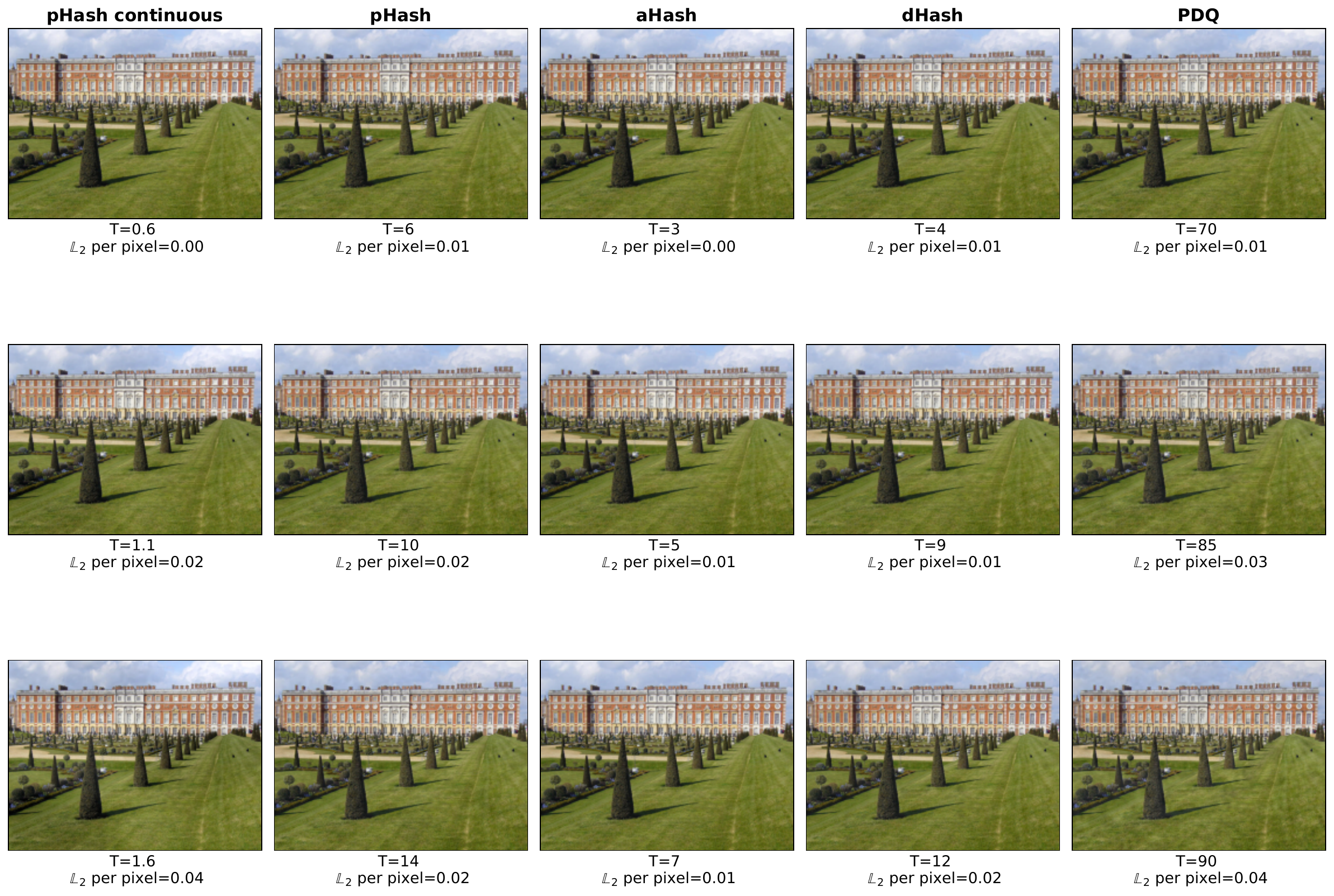}
    }
    \caption{Original image with modifications obtained using our black-box attack for different perceptual hashing algorithms and thresholds. We show the $\mathbb{L}_2$ perturbation per pixel. All the modified images are qualitatively very similar to the original image.}
    \label{fig:modified-images-3}  
    \end{figure*}
    
    \begin{figure*}[htbp]
    \centering
    \subfigure{
        \includegraphics[scale=0.4]{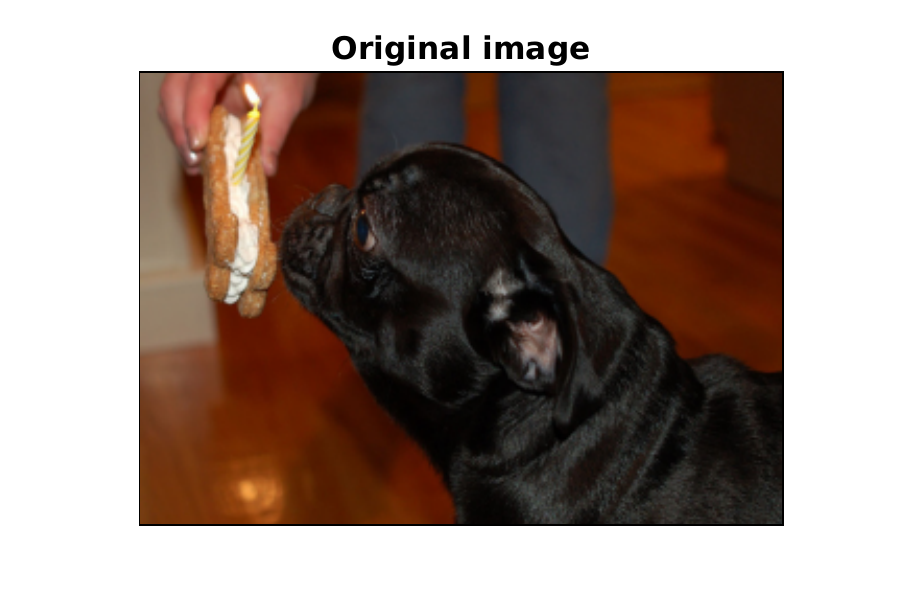}
    }
    \subfigure{
        \includegraphics[scale=0.42]{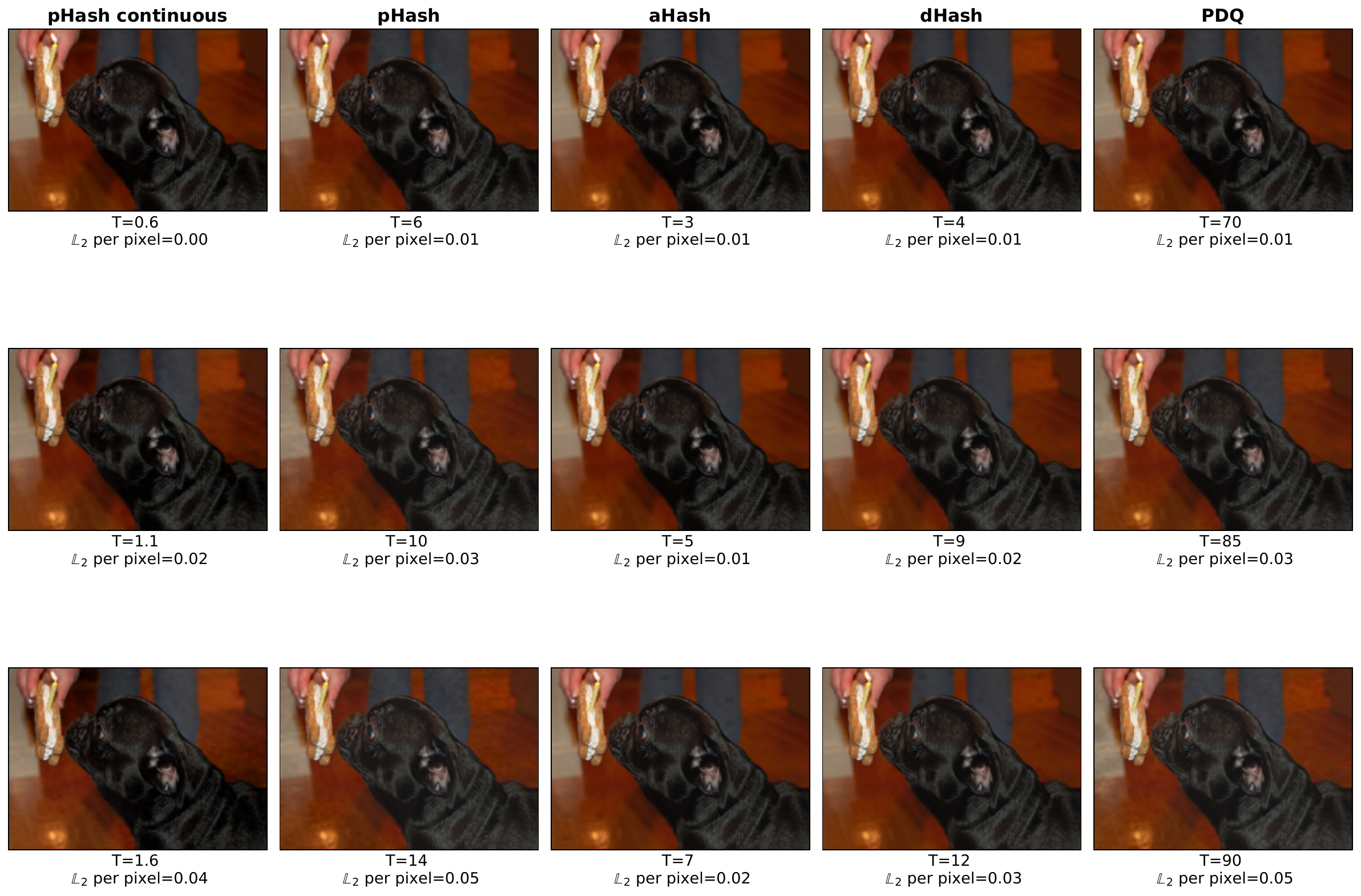}
    }
    \caption{Original image with modifications obtained using our black-box attack for different perceptual hashing algorithms and thresholds. We show the $\mathbb{L}_2$ perturbation per pixel. All the modified images are qualitatively very similar to the original image.}
    \label{fig:modified-images-4}  
    \end{figure*}
    
    \begin{figure*}[htbp]
    \centering
    \subfigure{
        \includegraphics[scale=0.4]{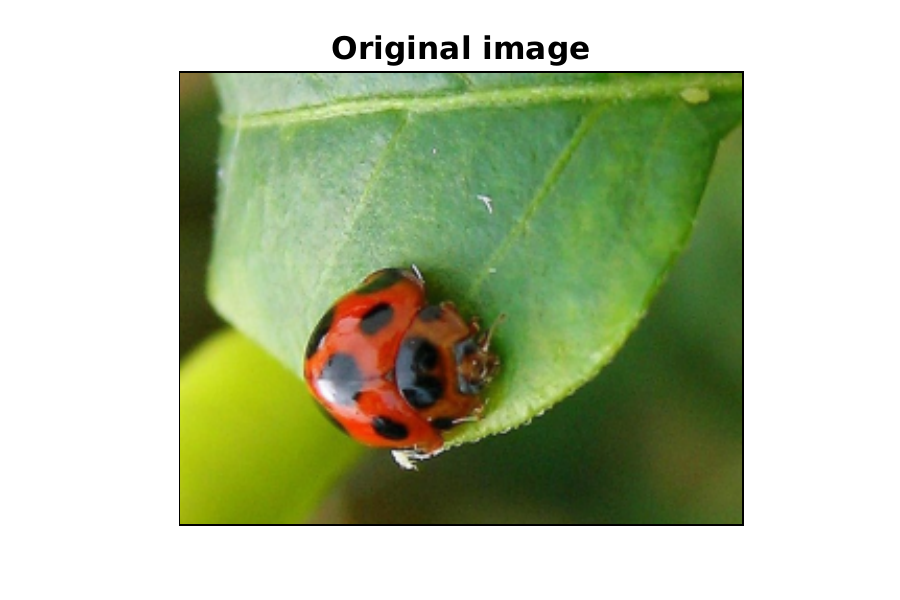}
    }
    \subfigure{
        \includegraphics[scale=0.42]{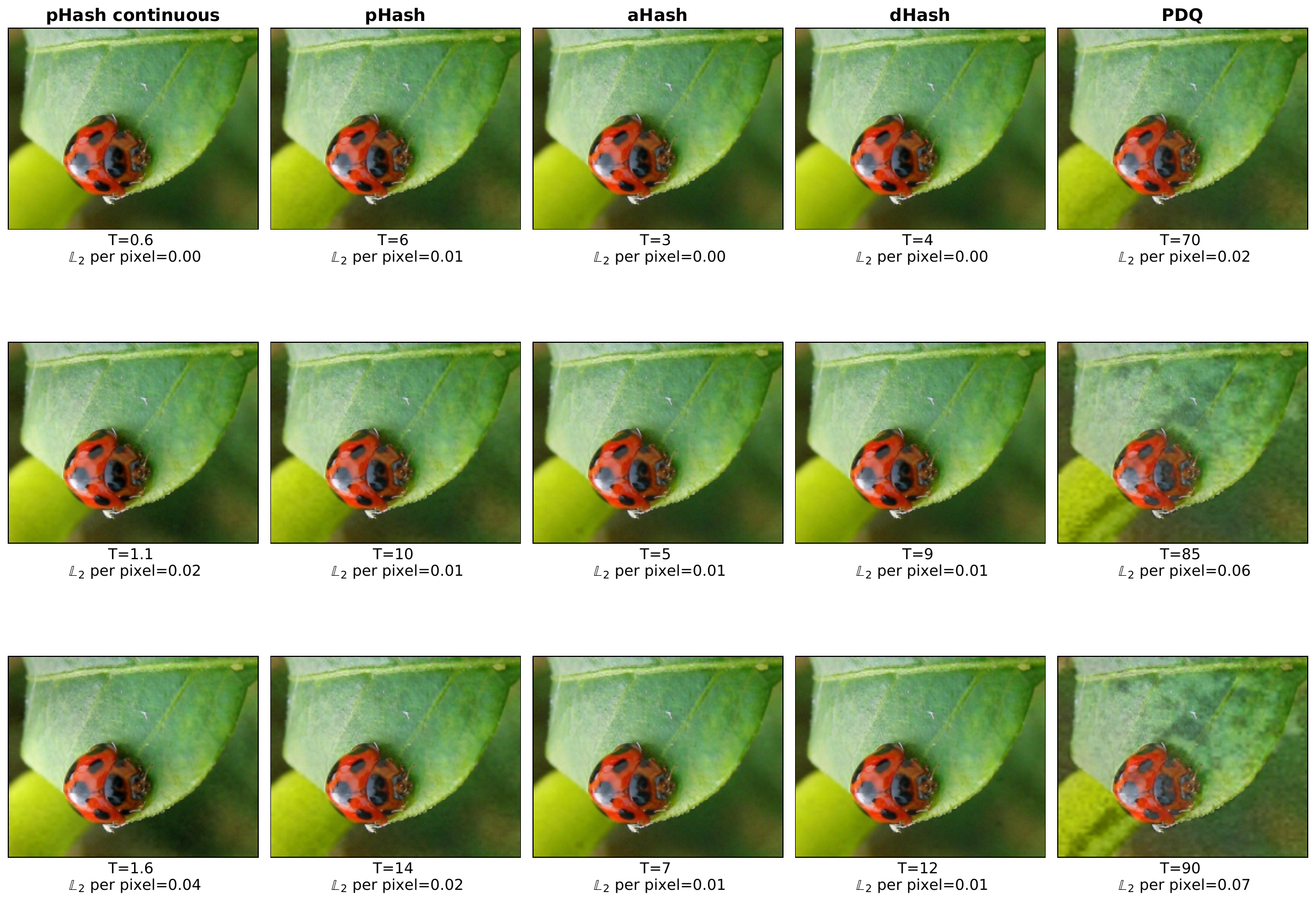}
    }
    \caption{Original image with modifications obtained using our black-box attack for different perceptual hashing algorithms and thresholds. We show the $\mathbb{L}_2$ perturbation per pixel. All the modified images are qualitatively very similar to the original image.}
    \label{fig:modified-images-5}  
    \end{figure*}
}

\section{Prevalence rate}
Table \ref{table:prevalence-small} shows the number of wrongly detected images for different perceptual hashing algorithms and threshold, for a prevalence rate of $10^{-7}$ for illegal content among 4.5B images shared daily. The results are comparable in order of magnitude to those of Table 2 in the main paper.

\begin{table}[h!]
\centering
\begin{tabular}{|cccc|} 
 \hline
 \multirow{2}{*}{\textbf{Hash}} & \multicolumn{3}{c|}{\textbf{Rough estimate of the number of}} \\
 & \multicolumn{3}{c|}{\textbf{wrongly detected images daily / }} \\
 & \multicolumn{3}{c|}{\textbf{Threshold $T$}} \\
 \hline
 pHash con- & $\sim18$M & $\sim179$M & $\sim 660$M\\
 tinuous  & $T=0.6$ & $T=1.1$ & $T=1.6$  \\ 
 \hline
  \multirow{2}{*}{pHash}   & $\sim 2$M & $\sim 104$M & $\sim3.3$B\\
   & $T=6$ & $T=10$ & $T=14$  \\ 
 \hline
   \multirow{2}{*}{aHash}  & $\sim 175$M & $\sim 496$M & $\sim1.1$B\\
   & $T=3$ & $T=5$ & $T=7$  \\ 
 \hline
   \multirow{2}{*}{dHash}   & $\sim3$M & $\sim 106$M & $\sim796$M\\
   & $T=4$ & $T=9$ & $T=12$  \\ 
 \hline
  \multirow{2}{*}{PDQ} & $\sim1$M & $\sim130$M & $\sim1.6$B\\
  & $T=70$ & $T=85$ & $T=90$  \\ 
 \hline
\end{tabular}
\caption{Rough estimate of the number of images that would be wrongly detected daily on WhatsApp alone for different perceptual hashing algorithms and thresholds. We here consider a database size of $N=100$k,  4.5B images being shared daily \cite{whatsappblog2017}, and a prevalence rate of illegal content of $10^{-7}$.}
\label{table:prevalence-small}
\end{table}

\section{Impact of diversity on visual similarity of perturbed images}

\begin{figure*}[h!]
\centering
\includegraphics[scale=0.4]{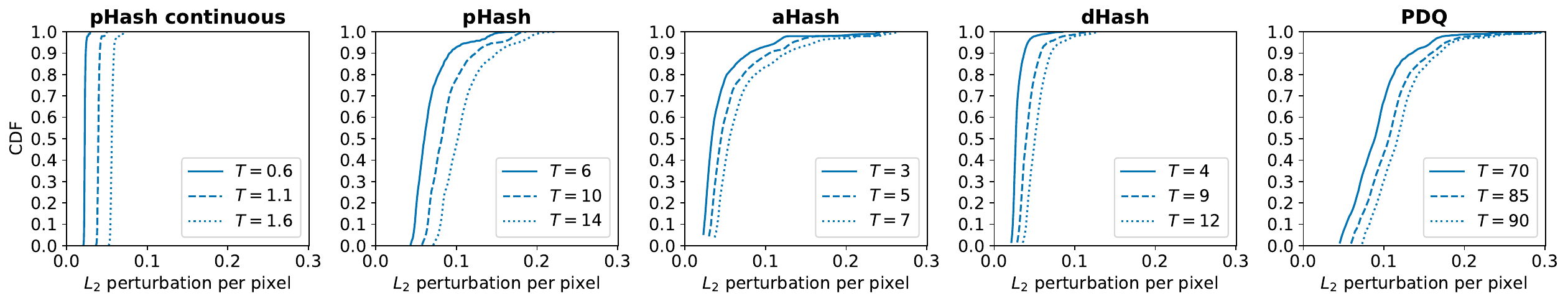}
\caption{Cumulative distribution function (CDF) for the $\mathbb{L}_2$ perturbations per pixel of the multiple perturbations generated for the same image for different algorithms and threshold. The number of perturbations generated for each image is $D = 50$ and the number of images attacked is $N' = 100$.}
\label{fig:diverse-perturbation-l2}  
\end{figure*}

Fig~\ref{fig:diverse-perturbation-l2} shows that our modifications to black-box algorithm leads to more diverse perturbations, but it also leads to relatively higher $\mathbb{L}_2$ perturbation per pixel compared to black-box algorithm without modifications. However we also observe that even for the highest thresholds, 100\% of diverse perturbed images generated for dHash (respectively 100.0\%, 84.5\%, 91.7\% and 72.7\% for pHash continuous, pHash, aHash and PDQ) of perturbed images are within the $\mathbb{L}_2$ perturbation per pixel of 0.13.

\section{Probability to flag $k$ images}

We consider a simple model with two types of users each sending $N$ images independently from one another. The first type, which we call non-offender, sends only non-illegal images. The second type, which we call offender, sends $l$ illegal images that are adversarially modified using our attack and $N-l$ non-illegal images. The probability for the detection system to flag an image is $FPR$ for non-illegal images and $1-FNR$ for illegal images adversarially modified using our attack.

Under this model, the number of images flagged for a non-offender follows a binomial distribution $N_1 \sim Bin(N, FPR)$. Similarly, the number of images flagged for an offender can be written as the sum of two binomial random variables $N_2 \sim Bin(l, 1 - FNR) + Bin(N-l, FPR)$. The two follow the same distribution if $1 - FNR = FPR$, i.e., if the adversarial images are indistinguishable from natural images to the detection system.

This allows us to compute, for each type of user, the probability that $0 \leq k \leq N$ of their images are flagged:

\begin{equation*}
\begin{aligned}
P(N_1=k) &= {N \choose k} FPR^k  (1-FPR)^{N-k} 
\end{aligned}
\end{equation*}

\begin{equation*}
\begin{aligned}
P(N_2=k) = \sum_{i=0}^{min(l, k)}&{l \choose i} (1-FNR)^i FNR^{l-i} \times\\ 
&{N-l \choose k-i} FPR^{k-i}  (1-FPR)^{N-l-k+i} 
\end{aligned}
\end{equation*}

\arxiv{
Fig.~\ref{fig:csp-at-least-k-all} shows the  results of our analysis for different perceptual hashing algorithms, for parameter values of $N=1000$ (number of images shared), $l=100$ (number of illegal images shared by an offender), and $FPR$ and $FNR$ values obtained against a database size of 100k. The probabilities to flag at least $k$ images are similar for offenders and non-offenders for all the perceptual hashing algorithms.

\begin{figure*}[htbp] 
\centering
\subfigure{%
\includegraphics[width=0.475\textwidth]{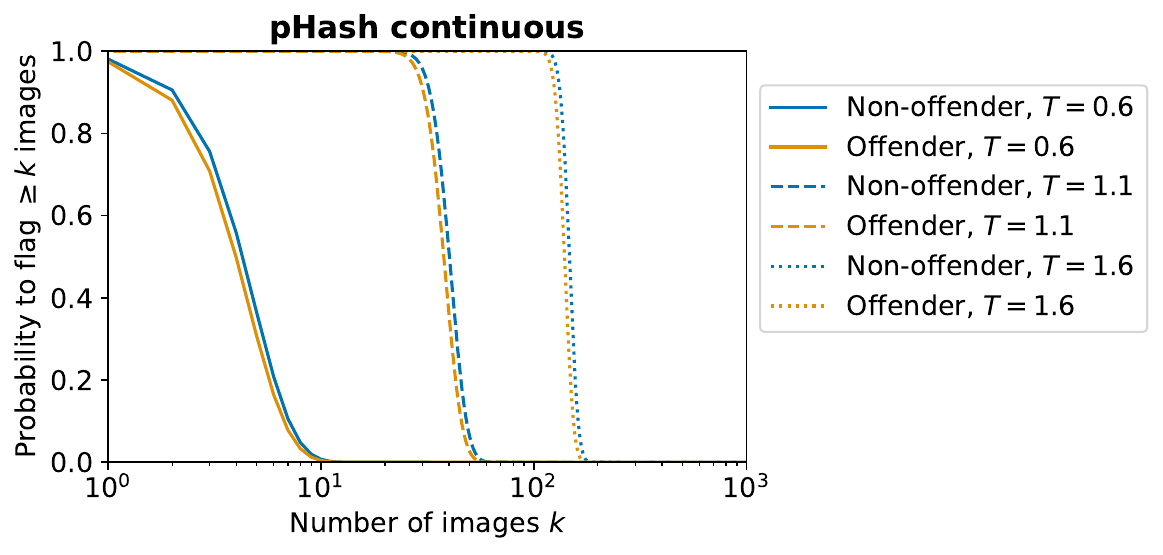}%
}\hfil
\subfigure{%
\includegraphics[width=0.475\textwidth]{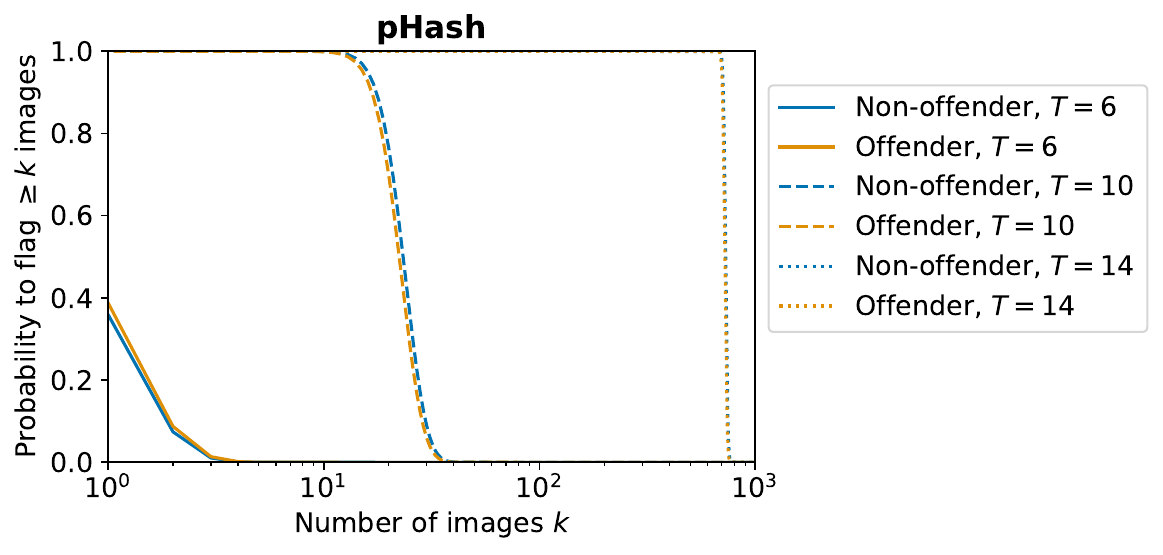}%
}

\subfigure{%
\includegraphics[width=0.475\textwidth]{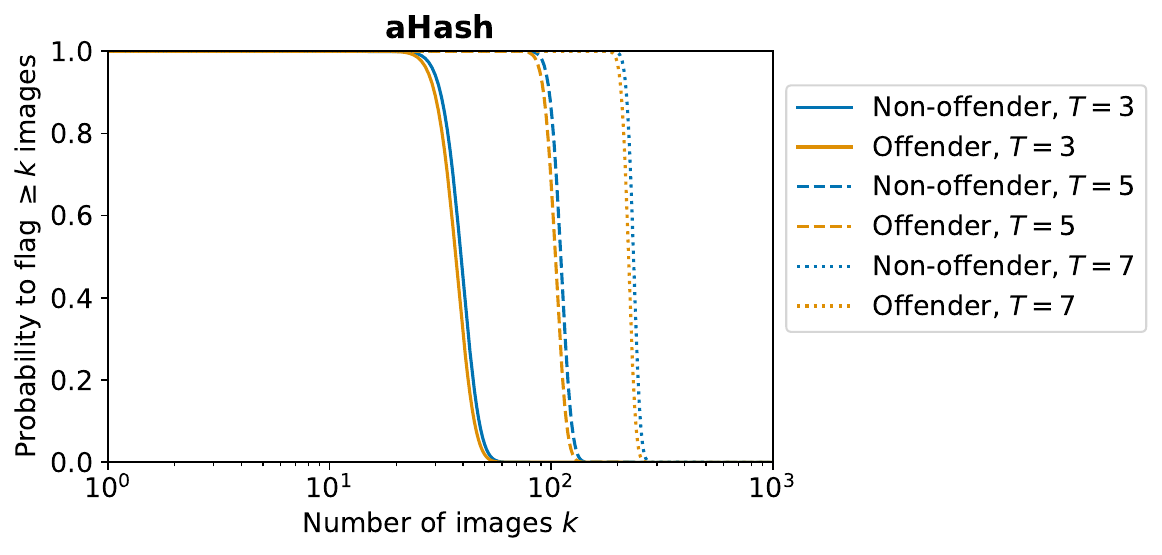}%
}\hfil
\subfigure{%
\includegraphics[width=0.475\textwidth]{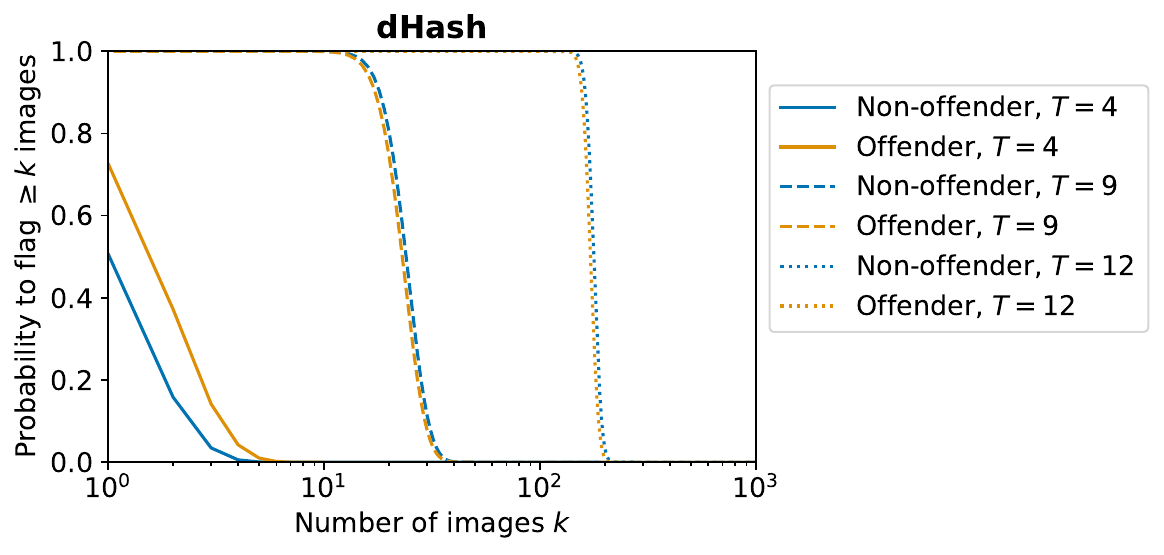}%
}

\caption{Probability that a detection system flags at least $k$ images shared by non-offenders (sharing 0 illegal images) and offenders (sharing 100 illegal images). Both offenders and non-offenders share a total of 1,000 images.}
\label{fig:csp-at-least-k-all}
\end{figure*}

}

\arxiv{
    \section{Distances between image hashes}
    
    Fig. \ref{fig:cdf-distances} shows the cumulative distribution function (CDF) for distances between the image hashes for different algorithms, number of images $P$ and datasets. We use the ImageNet and Stanford Dogs datasets. To compute each CDF, we randomly sample $P$ images from each dataset; for each hashing algorithm we compute the distances between image hashes for up to 1M pairs. For small values of $P$, we use all possible pairs while for larger $P$ we sample 1M random pairs.
    
    \begin{figure*}[htbp]
    \centering
    \includegraphics[scale=0.35]{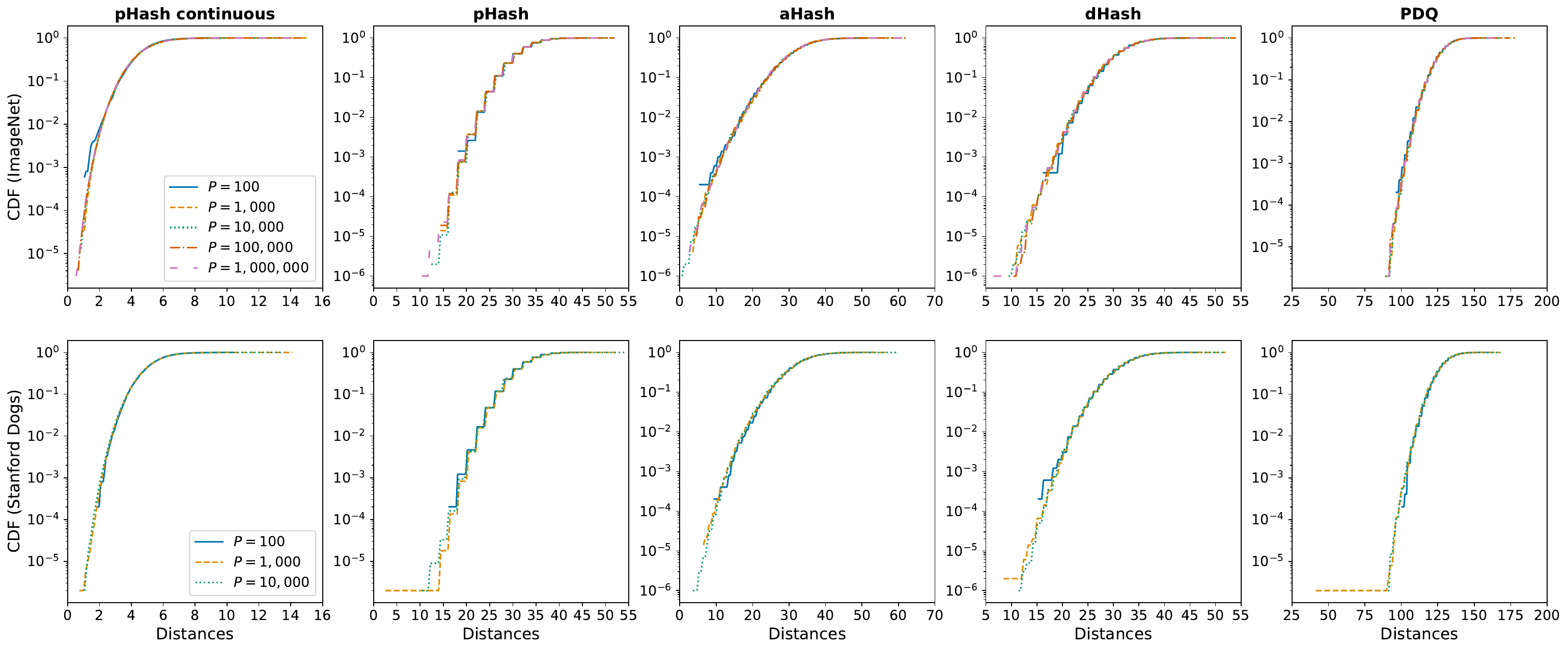}
    \caption{Cumulative distribution function (CDF) for distances between the image hashes for different algorithms, number of images $P$ and datasets. The CDFs are roughly stable for each algorithm for different number of images.}
    \label{fig:cdf-distances}  
    \end{figure*}
}

\end{document}